\documentclass{sig-alternate}
\usepackage{amssymb}
\usepackage{amsmath}
\usepackage{latexsym}
\usepackage{graphicx}
\usepackage{subfigure}
\usepackage{psfrag}
\usepackage{PageSetting}
\usepackage{wrapfig,epsfig}
\usepackage{algorithm}
\usepackage{algorithmic}
\newtheorem{remark}{Remark}
\newcommand{\btab}{\begin{tabbing}}
\newcommand{\etab}{\end{tabbing}}

\newcommand{\ppos}{iiiii\=iiiiiii\=iiiiiii\=iiiiiiiii\=iiii\=iiii\=iiii\=iiiii
\=iiiiiiiiiiiiiiiiiiii\=iiiiiiiiiiiiiiiii \=iiiiiiiii\=iiiiiii\kill}

\begin{document}

\title{Inferential or Differential: Privacy Laws Dictate}
 \author{Ke Wang, Peng Wang \\wangk@cs.sfu.ca,pwa22@sfu.ca\\Simon Fraser University
 \and Ada Waichee Fu \\adafu@cse.cuhk.edu.hk\\Chinese University of Hong Kong
 \and Raywong Chi-Wing Wong\\raywong@cse.ust.hk\\Hong Kong University of Science and
 Technology}
\maketitle

\begin{abstract}
So far, privacy models follow two paradigms. The first paradigm,
termed \emph{inferential privacy} in this paper, focuses on the risk
due to statistical inference of sensitive information about a target
record from other records in the database. The second paradigm,
known as \emph{differential privacy}, focuses on the risk to an
individual when included in, versus when not included in, the
database. The contribution of this paper consists of two parts. The
first part presents a critical analysis on differential privacy with
two results: (i) the differential privacy mechanism does not provide
inferential privacy, (ii) the impossibility result about achieving
Dalenius's privacy goal \cite{Dwork06} is based on an adversary
simulated by a Turing machine, but a \emph{human} adversary may
behave differently; consequently, the practical implication of the
impossibility result remains unclear. The second part of this work
is devoted to a solution addressing three major drawbacks in
previous approaches to inferential privacy: lack of flexibility for
handling variable sensitivity, poor utility, and vulnerability to
auxiliary information.

%
\end{abstract}

\section{Introduction}\label{introduction}
There has been a significant interest in the analysis of data sets
whose individual records are too sensitive to expose directly.
Examples include medical records, financial data, insurance data,
web query logs, user rating data for recommender systems, personal
data from social networks, etc. Data of this kind provide rich
information for data analysis in a variety of important
applications, but access to such data may pose a significant risk to
individual privacy, as illustrated in the following example.

\begin{example}\label{example1} A hospital maintains an online
database for answering count queries on medical data like the table
$T$ in Table \ref{tab:example1}. $T$ contains three columns, Gender,
Zipcode, and Disease, where Disease is a sensitive attribute.
Suppose that an adversary tries to infer the disease of an
individual Alice, with the background knowledge that Alice, a female
living in the area with Zipcode 61434, has a record in $T$. The
adversary issues the following two queries $Q_1$ and $Q_2$:

$Q_1$: SELECT COUNT(*) FROM T WHERE Gender=F AND Zipcode=61434

$Q_2$: SELECT COUNT(*) FROM T WHERE Gender=F AND Zipcode=61434 AND
Disease=HIV

Each query returns the number of participants (records) who match
the description in the WHERE clause. Suppose that the answers for
$Q_1$ and $Q_2$ are $x$ and $y$, respectively. The adversary then
estimates that Alice has HIV with probability $y/x$, and if $y/x$
and $x$ are ``sufficiently large", there will be a privacy breach.


\end{example}

\begin{table}
\centering
\begin{scriptsize}
\begin{tabular}{| c | c | c|} \hline
 Gender & Zipcode & Disease \\ \hline
 M & 54321 & Brain Tumor \\ \hline
 M & 54322 & Indigestion \\ \hline
 F & 61234 & Cancer \\ \hline
 F & 61434 & HIV \\ \hline
 ... & ... & ... \\ \hline
\end{tabular}
\caption{A table $T$} \label{tab:example1}
\end{scriptsize}
\end{table}

\subsection{Inferential vs Differential}
In the above example, the adversary infers that the rule
\[ (Gender=F \wedge Zipcode=61434) \rightarrow (Disease=HIV) \]
holds with the probability $y/x$ and that Alice has HIV with the
probability $y/x$, assuming that the (diseases of) records follow
some underlying probability distribution. This type of reasoning,
which learns information about \emph{one record} from the statistics
of \emph{other records}, is found in many advanced applications such
as recommender systems, prediction models, viral marketing, social
tagging, and social networks. The same technique could be misused to
infer sensitive information about an individual like in the above
example. According to the \emph{Privacy Act} of Canada, publishing
the above query answers would breach Alice's privacy because they
disclose Alice's disease with a high accuracy. In this paper,
\emph{inferential privacy} refers to the requirement of limiting the
statistical inference of sensitive information about a target record
from other records in the database. See \cite{Adam89} for a list of
works in this field.

One recent breakthrough in the study of privacy preservation is
\emph{differential privacy}
\cite{Dwork06}\cite{springerlink:10.1007/11681878.14}. In an
``impossibility result", the authors of
\cite{Dwork06}\cite{springerlink:10.1007/11681878.14} showed that it
is impossible to achieve Dalenius's absolute privacy goal for
statistical databases: anything that can be learned about a
respondent from the statistical database should be learnable without
access to the database. Instead of limiting what can be learnt about
one record from other records, the differential privacy mechanism
hides the presence or absence of a participant in the database,

by producing noisy query answers such that the distribution of query
answers changes very little when the database differs in any
\emph{single} record. The following definition is from
\cite{Blum08}.

\begin{definition}
A randomized function $K$ gives $\varepsilon$-differential privacy
if for all data sets $T$ and $T'$ differing on at most one record,
for all queries $Q$, and for all outputs $x$, $Pr[K(T,Q) = x] \leq
exp(\varepsilon) Pr[K(T',Q) = x]$.
\end{definition}

With a small $\varepsilon$, the presence or absence of an individual
is hidden because $T$ and $T'$ are almost equally likely to be the
underlying database that produces the final output of the query.
Some frequently cited claims of the differential privacy mechanism
are that it provides privacy without any assumptions about the data
and that it protects against arbitrary background information. But
there is no free lunch in data privacy, as pointed out by Kifer and
Machanavajjhala recently \cite{Kifer2011}. Their study shows that
assumptions about the data and the adversaries are required if
hiding the \emph{evidence} of participation, instead of the
presence/absence of records in the database, is the privacy goal,
which they argue should be a major privacy definition.

\subsection{Contributions}

The contribution of this paper consists of two parts.
In the first part, we argue that differential privacy is
insufficient because it does not provide inferential privacy. We
present two specific results:

\begin{itemize}
\item (Section 2.1) Using a \emph{differential inference theorem}, we show that the noisy query answers returned by
the differential privacy mechanism may derive an inference
probability that is arbitrarily close to the inference probability
obtained from the noise-free query answers. This study suggests that
providing inferential privacy remains a meaningful research problem,
despite the protection of differential privacy.

\item (Section 2.2) While the impossibility result in \cite{Dwork06} is based on an adversary
simulated by a Turing machine, a \emph{human} adversary may behave
differently when evaluating the sensitivity of information. We use
the Terry Gross example, which is a key motivation of differential
privacy, to explain this point. This study suggests that the
practical implication of the impossibility result remains unclear.

\end{itemize}

Given that inferential privacy remains relevant, the second part of
this work is devoted to stronger solutions for inferential privacy.
Previous approaches suffer from three major limitations. Firstly,
most solutions are unable to handle sensitive values that have
skewed distribution and varied sensitivity. For example, with the
Occupation attribute in the Census data (Section 7) having the
minimum and maximum frequency of 0.18\% and 7.5\%, the maximum
$\ell$-diversity \cite{MKG+006} that can be provided is
$13$-diversity because of the eligibility requirement $1/\ell \geq
7.5\%$ \cite{XT06b}. Therefore, it is impossible to protect the
infrequent items at the tail of the distribution or more sensitive
items by a larger $\ell$-diversity, say $50$-diversity, which is
more than 10 times the prior 0.18\%. Secondly, even if it is
possible to achieve such $\ell$-diversity, enforcing
$\ell$-diversity with a large $\ell$ across \emph{all} sensitive
values leads to a large information loss. Finally, previous
solutions are vulnerable to additional auxiliary information
\cite{TaoCorr08}\cite{Kifer:2009:APD:1559845.1559861}\cite{Li09}. We
address these issues in three steps.

\begin{itemize}

\item (Section 3) To address the first two limitations in the above,
we consider a sensitive attribute with domain values $x_1,\cdots,
x_m$ such that each $x_i$ has a different sensitivity, thus, a
tolerance $f'_i$ on inference probability. We consider a
bucketization problem in which buckets of \emph{different} sizes can
be formed to accommodate different requirements $f'_i$. The goal is
to find a collection of buckets for a given set of records so that a
notion of information loss related to bucket size is minimized and
the privacy constraint $f'_i$ of all $x_i$'s is satisfied.

\item (Sections 4, 5, and 7)

We present an efficient algorithm for the case of two distinct
bucket sizes (but many buckets) with \emph{guaranteed optimality},
and a heuristic algorithm for the general case. The empirical study
on real life data sets shows that both solutions are good
approximations of optimal solutions in the general case and better
deal with a sensitive attribute of skewed distribution and varied
sensitivity.

 \item (Section 6) We adapt our solutions to guard against two previously identified strong attacks,
 corruption attack \cite{TaoCorr08} and negative association attack
 \cite{Kifer:2009:APD:1559845.1559861}\cite{Li09} (see more details in Section 6).

\end{itemize}

\subsection{Related Work}

Limiting statistical disclosure has been a topic extensively studied
in the field of statistical databases, see \cite{Adam89} for a list
of works. This problem was recently examined in the context of
privacy preserving data publishing and some representative privacy
models include $\rho_1$-$\rho_2$ privacy
\cite{Evfimievski:2003:LPB:773153.773174}, $\ell$-diversity
principle \cite{MKG+006}, and $t$-closeness\cite{N07}. All of these
works assume uniform sensitivity across all sensitive values. One
exception is the personalized privacy in \cite{XT06a} where a record
owner can specify his/her privacy threshold. Another exception is
\cite{LW10} where each sensitive value may have a different privacy
setting. To achieve the privacy goal, these works require a taxonomy
of domain values to generalize the attributes, thus, cannot be
applied if such taxonomy is not available. The study in \cite{XT06b}
shows that generalized attributes are not useful for count queries
on raw values. Dealing with auxiliary information is a hard problem
in data privacy
\cite{TaoCorr08}\cite{Kifer:2009:APD:1559845.1559861}\cite{Li09},
and so far there is little satisfactory solution.

There have been a great deal of works in differential privacy since
the pioneer work
\cite{springerlink:10.1007/11681878.14}\cite{Dwork06}. This
includes, among others, contingency table releases
\cite{Barak:2007:PAC:1265530.1265569}, estimating the degree
distribution of social networks \cite{Hay09}, histogram queries
\cite{Hay10} and the number of permissible queries \cite{XT08}.
These works are concerned with applications of differential privacy
in various scenarios. Unlike previous works, the authors of
\cite{Kifer2011} argue that hiding the evidence of participation,
instead of the presence/absence of records in the database, should
be a major privacy definition, and this privacy goal cannot be
achieved with making assumptions about the data and the adversaries.

\section{Analyzing Differential Privacy}\label{differential}

This section presents a critical analysis on the differential
privacy mechanism. In Section \ref{independent} we show that the
differential privacy mechanism allows violation of inferential
privacy. In Section \ref{Gross} we argue that a human adversary may
behave differently from some assumptions made in the impossibility
result of \cite{Dwork06}, thus, the practical implication of the
impossibility result remains unclear.

\subsection{On Violating Inferential Privacy}\label{independent}

One popularized claim of the differential privacy mechanism is that
it protects an individual's information even if an attacker knows
about all other individuals in the data. We quote the original
discussion from \cite{Blum:2005:PPS:1065167.1065184} (pp 3):

\begin{quote}
``If there is information about a row that can be learned from other
rows, this information is not truly under the control of that row.
Even if the row in question were to sequester itself away in a high
mountaintop cave, information about the row that can be gained from
the analysis of other rows is still available to an adversary. It is
for this reason that we focus our attention on those inferences that
can be made about rows without the help of others."

\end{quote}

In other words, the differential privacy framework does not consider
violation to inferential privacy and the reason is that it is not
under the control of the target row. Two points need clarification.
Firstly, a user submits her sensitive data to an organization
because she trusts that the organization will do everything possible
to protect her sensitive information; indeed, the data publisher has
full control in how to release the data or query answers in order to
protect individual privacy. Secondly, learning information about one
record from other records could pose a risk to an individual if the
learnt information is accurate about the individual. This type of
learning assumes that records follow some underlying probability
distribution, which is widely adapted by prediction models in many
real applications. Under this assumption, suppose $Q_1$ and $Q_2$ in
Example \ref{example1} have the answers $x=100$ and $y=99$,
even if Alice's record is removed from the database, it is still
valid to infer that Alice has HIV with a high probability.


Next, we show that even if the differential privacy mechanism adds
noises to the answers for queries $Q_1$ and $Q_2$, Alice's disease
can still be inferred using the noisy answers.

Let $x$ and $y$ be the true answers to $Q_1$ and $Q_2$. We assume
that $x$ and $y$ are non-zero.
The differential privacy mechanism will return the noisy answers
$X=x+\xi_1$ and $Y=y+\xi_2$ for $Q_1$ and $Q_2$, after adding noises
$\xi_1$ and $\xi_2$. Consider the most used Laplace distribution
$Lap(b)=\frac{1}{2b} exp(-|\xi|/b)$ for the noise $\xi$, where $b$
is the scale factor. The mean $E[\xi]$ is zero and the variance
$var[\xi]$ is $2b^2$. The next theorem  is due to \cite{Dwork06}.

\begin{theorem}\cite{Dwork06} \label{th:countquery} For a count query $Q$, the mechanism $K$ that adds
independently generated noise $\xi$ with distribution
$Lap(1/\varepsilon)$ to the output enjoys $\varepsilon$-differential
privacy.
\end{theorem}

The next theorem shows that $Y/X$ is a good approximation of $y/x$.

\begin{theorem}[Differential Inference Theorem]  \label{variance}
Given two queries  $Q_1$ and $Q_2$ as above, let $x$ and $y$ be the
true answers and let $X$ and $Y$ be the answers returned by the
$\varepsilon$-differential privacy mechanism.
$E[\frac{Y}{X}]=\frac{y}{x}(1+\frac{2b^2}{x^2})$ and
$var[\frac{Y}{X}]=\frac{2b^2}{x^2}(1+(\frac{y}{x})^2)$, where
$b=1/\varepsilon$.

%
\end{theorem}

\begin{proof} Using the Taylor expansion technique  \cite{Johnson80} \cite{Stuart98}, the mean
$E[\frac{Y}{X}]$ and variance $var[\frac{Y}{X}]$ of $Y/X$ can be
approximated as follows:
\[
E[\frac{Y}{X}]\simeq \frac{E[Y]}{E[X]}+\frac{cov[X,Y]}{E[X]^2} +
\frac{var[X]E[Y]}{E[X]^3} \]
\[
var[\frac{Y}{X}]\simeq \frac{var[Y]}{E[X]^2} - \frac{2E[Y]}{E[X]^3}
cov[X,Y] + \frac{E[Y]^2}{E[X]^4} var[X] \label{var}
\]
$E[X]$ and $E[Y]$ are equal to the true answers $x$ and $y$ of $Q_1$
and $Q_2$. $var[X]$ and $var[Y]$ are $2b^2$ for $Lap(b)$.
$cov[X,Y]=cov[x+\xi_1,y+\xi_2]=cov[\xi_1,\xi_2]$. Since $\xi_1$ and
$\xi_2$ are unrelated, $cov[\xi_1,\xi_2]=0$. Simplifying the above
equations, we get $E[\frac{Y}{X}]$ and $var[\frac{Y}{X}]$ as
required.

\end{proof}

The next corollary follows from the fact that $\frac{y}{x}\leq 1$
and $b$ is a constant for a given $\varepsilon$-differential privacy
mechanism $K$.

\begin{corollary}\label{c1}
Let $X,Y$ be defined as in Theorem \ref{variance}. As the query size
$x$ for $Q_1$ increases, $E[\frac{Y}{X}]$ gets arbitrarily close to
$\frac{y}{x}$ and $var[\frac{Y}{X}]$ gets arbitrarily close to zero.
\end{corollary}

Corollary \ref{c1} suggests that $Y/X$, where $Y$ and $X$ are the
noisy query answers returned by the differential privacy mechanism,
can be a good estimate of the inference probability $y/x$ for a
large query answer $x$. For example, for $\varepsilon=0.1$ and
$x=100$, $\frac{2b^2}{x^2}=0.02$, and following Theorem
\ref{variance}, $E[\frac{Y}{X}]$ is 1.02 times $\frac{y}{x}$; if
$x=1000$, $E[\frac{Y}{X}]$ is 1.0002 times $\frac{y}{x}$. If $y/x$
is high, inferential privacy is violated. Note that
$var[\frac{Y}{X}]$ is small in these cases.

\subsection{On The Impossibility Results}\label{Gross}

A key motivation behind differential privacy is the impossibility
result about the Dalenius's privacy goal  \cite{Dwork06}.
Intuitively, it says that for any privacy mechanism and any
distribution satisfying certain conditions, there is always some
particular piece of auxiliary information, $z$, so that $z$ alone is
useless to an adversary who tries to win, while $z$ in combination
with access to the data through the privacy mechanism permits the
adversary to win with probability arbitrarily close to 1. The proof
assumes an adversary simulated by a Turing machine. We argue that a
\emph{human} adversary, who also considers the ``semantics" when
evaluating the usefulness of information, may behave differently.
Let us explain this point by the Terry Gross example that was
originally used to capture the intuition of the impossibility result
in \cite{Dwork08}.

In the Terry Gross example, the exact height is considered private,
thus, useful to an adversary, whereas the auxiliary information of
being two inches shorter than an unknown average is considered not
private, thus, not useful. Under this assumption, accessing the
statistical database, which returns the average height, is to blame
for disclosing Terry Gross's privacy. Mathematically, knowing the
exact height is a remarkable progress from knowing two inches
shorter than an unknown average. However, to a \emph{human}
adversary, the information about how an individual \emph{deviates
from the statistics} already discloses the sensitive information,
regardless of what the statistics is. For example, once knowing that
someone took the HIV check-up ten times more frequently than an
unknown average, his/her privacy is already leaked. Here, a human
adversary is able to interpret ``deviation" as a sensitive notion
based on ``life experiences", even though mathematically deviation
does not derive the exact height. It is unclear whether such a human
adversary can be simulated by a Turing machine.

In practice, a realistic privacy definition does allow disclosure of
sensitive information in a controlled manner and there are scenarios
where it is possible to protect inferential privacy while retaining
a reasonable level of data utility. For example, the study in
\cite{FWY05} shows that the anonymized data is useful for training a
classifier because the training does not depend on detailed personal
information. Another scenario is when the utility metric is
different from the adversary's target. Suppose that the attribute
$Disease$ is sensitive and the response attribute $R$ (to a
medicine) is not. Learning the following rules does not violate
privacy
\begin{center} $(Disease=x_1) \rightarrow
(R=Positive)$\\
$(Disease=x_2) \rightarrow (R=Positive)$ \end{center} in that a
positive response does not indicate a specific disease with
certainty. However, these rules are useful for a researcher to
exclude the diseases $x_1$ and $x_2$ in the absence of a positive
response. Even for a sensitive attribute like $Disease$, the varied
sensitivity of domain values (such as Flu and HIV) could be
leveraged to retain more utility for less sensitive values while
ensuring strong protection for highly sensitive items. In the rest
of the paper, we present an approach of leveraging such varied
sensitivity to address some drawbacks in previous approaches to
inferential privacy.

 \section{Problem Statement}\label{problem}
This section defines the problem we will study. First, we present
our model of adversaries, privacy, and data utility.

\subsection{Preliminaries}
The database is a microdata table $T(QI,SA)$ with each record
corresponding to a participant. $QI$ is a set of non-sensitive
attributes $\{A_1,\cdots,A_d\}$. $SA$ is a sensitive attribute and
has the domain $\{x_1,\cdots,x_m\}$. $m$ is the domain size of $SA$,
also written $|SA|$. Each $x_i$ is called a sensitive value or a
$SA$ value. $o_i$ denotes the number of records for $x_i$ in $T$ and
$f_i$ denotes the frequency $o_i/|T|$, where $|T|$ is the
cardinality of $T$. For a record $r$ in $T$, $t[QI]$ and $r[SA]$
denote the values of $r$ on $QI$ and $SA$. Table \ref{tab1} lists
some of the notations used in this paper.

An adversary wants to infer the $SA$ value of a target individual
$t$. The adversary has access to a published version of $T$, denoted
by $T^*$. For each $SA$ value $x_i$, $Pr(x_i|t,T^*)$ denotes the
probability that $t$ is inferred to have $x_i$. For now, we consider
an adversary with the following auxiliary information: a $t$'s
record is contained in $T$, $t$'s values on $QI$, i.e., $t[QI]$, and
the algorithm used to produce $T^*$. Additional auxiliary
information will be considered in Section \ref{extension}.

One approach for limiting $Pr(x_i|t,T^*)$ is \emph{bucketization}
\cite{XT06b}. In this approach, the records in $T$ are grouped into
small-size buckets and each bucket is identified by a unique bucket
ID, $BID$. We use $g$ to refer to both a bucket and the bucket ID of
a bucket, depending on the context. $T^*$ is published in two
tables, $QIT(QI,BID)$ and $ST(BID,SA)$. For each record $r$ in $T$
that is grouped into a bucket $g$, $QIT$ contains a record
$(r[QI],g)$ and $ST$ contains a record $(g,r[SA])$ (with duplicates
preserved). For a target individual $t$ with $t[QI]$ contained in a
bucket $g$, the probability of inferring a $SA$ value $x_i$ using
$g$, $Pr(x_i|t,g)$, is equal to $|g,x_i|/|g|$, where $|g,x_i|$
denotes the number of occurrence of $(g,x_i)$ in $ST$ and $|g|$
denotes the size of $g$. $Pr(x_i|t,T^*)$ is defined to be the
maximum $Pr(x_i|t,g)$ for any bucket $g$ containing $t[QI]$
\cite{XT06b}.

\begin{example} For the microdata $T$ in Table~\ref{tab:example1},
Gender and Zipcode are the QI attributes and Disease is SA.
Table~\ref{tab:example-anonymized} shows the $QIT$ and $ST$ for one
bucketization. To infer the $SA$ value of Alice with $QI=\langle
F,61434\rangle$, the adversary first locates the bucket that
contains $\langle F,61434\rangle$, i.e., $BID=2$. There are two
diseases in this bucekt, Cancer and HIV, each occurring once. So
$Pr(x_i|Alice,2)=50\%$, where $x_i$ is either Cancer or HIV.
\end{example}

\begin{table}
\centering
\begin{scriptsize}
\begin{tabular}{c c}
\begin{minipage}[htbp]{3.5cm}
\center
\begin{tabular}{| c | c | c |}\hline
 Gender & Zipcode & BID \\ \hline
 M & 54321 & 1 \\ \hline
 M & 54322 & 1 \\ \hline
 F & 61234 & 2 \\ \hline
 F & 61434 & 2 \\ \hline
 ... & ... & ... \\ \hline
\end{tabular}
\end{minipage}
& \hspace*{0mm}
\begin{minipage}[htbp]{3cm}
\center
\begin{tabular}{| c | c |}\hline
 BID & Disease \\ \hline
 1 & Brain Tumor \\ \hline
 1 & Indigestion \\ \hline
 2 & Cancer \\ \hline
 2 & HIV \\ \hline
 ... & ... \\ \hline
\end{tabular}
\end{minipage}
\\ &
\\
(a) $QIT$ &   (b) $ST$
\end{tabular}
\caption{An anonymized table $T^*$} \label{tab:example-anonymized}
\end{scriptsize}
\end{table}

\begin{table}
\begin{tabular}{| c | c |} \hline
$T$, $|T|$ & \parbox[t]{6cm}{the raw data and its cardinality} \\
\hline
 $m$ & \parbox[t]{6cm}{domain size of $SA$} \\ \hline
 $x_i$ & \parbox[t]{6cm}{a sensitive value} \\ \hline
 $o_i$ & \parbox[t]{6cm}{number of occurrence of $x_i$ in $T$} \\ \hline
 $f_i$ & \parbox[t]{6cm}{$o_i/|T|$} \\ \hline
 $f'_i$ & \parbox[t]{6cm}{privacy threshold for $x_i$} \\ \hline
 $F'$-privacy & \parbox[t]{6cm}{a collection of $f'_i$ for $x_i$} \\ \hline
 $B_j(S_j,b_j)$ & \parbox[t]{6cm}{$b_j$ buckets of size $S_j$} \\ \hline
 $s(B_j)$ & \parbox[t]{6cm}{total size of buckets in $B_j$} \\ \hline
 \end{tabular}
\caption{Notations} \label{tab1}
\end{table}

\subsection{Privacy Specification}

We consider the following privacy specification.

\begin{definition}[$F'$-Privacy] For each $SA$ value $x_i$,
\emph{$f'_i$-privacy} specifies the requirement that
$Pr(x_i|t,T^*)\leq f'_i$, where $f'_i$ is a real number in the range
(0,1]. \emph{$F'$-privacy} is a collection of $f'_i$-privacy for all
$SA$ values $x_i$.
\end{definition}

For example, the publisher may set $f'_i=1$ for some $x_i$'s that
are not sensitive at all, set $f'_i$ manually to a small value for a
few highly sensitive values $x_i$, and set $f'_i=min\{1,a\times f_i+
b\}$ for the rest of $SA$ values whose sensitivity grows linearly
with their frequency, where $a$ and $b$ are constants. Our approach
assumes that $f'_i$ is specified but does not depend on how $f'_i$
is specified. The next lemma follows easily and the proof is
omitted.

\begin{lemma}\label{eligibility}
A bucketization $T^*$ satisfying $F'$-privacy exists if and only if
$f'_i \geq f_i$ for all $x_i$.
\end{lemma}

\begin{remark} \label{remark1} To model a
given $F'$-privacy specification by $\ell$-diversity \cite{MKG+006},
the smallest $\ell$ required is set by $\ell=\lceil 1/min_i f'_i
\rceil$. If some $x_i$ is highly sensitive, i.e., has a small
$f'_i$, this $\ell$ will be too large for less sensitive $x_i$'s.
This leads to poor utility for two reasons. First, the previous
bucketization \cite{XT06b} produces buckets of the sizes $\ell$ or
$\ell+1$. Thus, a large $\ell$ leads to large buckets and a large
information loss. Second, a large $\ell$ implies that the
\emph{eligibility requirement} \cite{XT06b} for having a
$\ell$-diversity $T^*$, i.e., $1/\ell \geq max_i f_i$, is more
difficult to satisfy. In contrast, the corresponding eligibility
requirement for having $F'$-privacy $T^*$ is $f'_i\geq f_i$ for all
$x_i$'s (Lemma \ref{eligibility}), which is much easier to satisfy.
In Section 3.4, we will address the large bucket size issue by
allowing buckets of different sizes to be formed to accommodate
different requirements $f'_i$.

\end{remark}

\subsection{Utility Metrics}
Within each bucket $g$, the $QI$ value of every record is equally
likely associated with the $SA$ value of every record through the
common $BID$. Therefore, the bucket size $|g|$ serves as a measure
of the ``disorder" of such association. This observation motivates
the following notion of information loss.

\begin{definition}\label{MSE}
Let $T^*$ consist of a set of buckets $\{g_1,\cdots,g_b\}$. The Mean
Squared Error (MSE) of $T^*$ is defined by
\begin{equation}
MSE(T^*)= \frac{\sum_{i=1}^b (|g_i|-1)^2}{|T|-1}
\end{equation}
\end{definition}

Any bucketization $T^*$ has a $MSE$ in the range $[0,|T|- 1]$. The
raw data $T$ is one extreme where each record itself is a bucket, so
$MSE=0$. The single bucket containing all records is the other
extreme where $MSE=|T|-1$.  With $|T|$ being fixed, to minimize
$MSE$, we shall minimize the following loss metric:
\begin{equation}
Loss(T^*)= \sum_{i=1}^b(|g_i|-1)^2 \label{loss}
\end{equation}
Note that $Loss$ has the additivity property: if $T^* = T_1^* \cup
T_2^*$, then $Loss(T^*)=Loss(T_1^*) + Loss(T_2^*)$.

\subsection{Problem Description}

To minimize $Loss$, we consider a general form of bucketization in
which buckets of different sizes can be formed so that a large
bucket size is used for records having a more sensitive $x_i$ (i.e.,
a small $f'_i$) and a small bucket size is used for records having
less sensitive $x_i$ (i.e., a larger $f'_i$). A collection of
buckets can be specified by a \emph{bucket setting} of the form
$\langle B_1(S_1,b_1),\cdots,B_q(S_q,b_q)\rangle$, where $b_j$ is
the number of buckets of the size $S_j$, $j=1,\cdots,q$, and
$S_1<\cdots < S_q$. We also denote a bucket setting simply by $\cup
B_j$. $s(B_j)=b_j S_j$ denotes the total size of the buckets in
$B_j$. Following Definition \ref{loss}, the collection of buckets
specified by $\cup B_j$ has the loss $\sum_{j=1}^q b_j\times
(S_j-1)^2$. We denote this loss by $Loss(\cup B_j)$.

A bucket setting $\cup B_j$ is \emph{feasible} wrt $T$ if $\sum_j
s(B_j)=|T|$. A feasible bucket setting is \emph{valid} wrt
$F'$-privacy if there is an assignment of the records in $T$ to the
buckets in $\cup B_j$ such that no $SA$ value $x_i$ has a frequency
more than $f'_i$ in any bucket $g$, i.e., $Pr(x_i|t,g)\leq f'_i$.
Such assignment is called a \emph{valid} record assignment.

\begin{definition}[Optimal multi-size bucket setting]\label{d2}
Given $T$ and $F'$-privacy, we want to find a valid bucket setting
$\langle B_1(S_1,b_1),\cdots,B_q(S_q,b_q)\rangle$ that has the
minimum $Loss(\cup B_j)$ among all valid bucket settings.
\end{definition}

This problem must determine the number $q$ of distinct bucket sizes,
each bucket size $S_j$ and the number $b_j$ of buckets for the size
$S_j$, $1\leq j\leq q$. The following special case is a building
block of our solution.

\begin{definition}[Optimal two-size bucket setting]\label{d22}
Given $T$ and $F'$-privacy, we want to find a valid two-size bucket
setting $\langle B_1 (S_1,b_1),B_2(S_2,b_2)\rangle$ that has the
minimum loss among all valid two-size bucket settings.
\end{definition}

\begin{remark}\label{remark2}
The bucket setting problem is challenging for several reasons.
Firstly, allowing varied sensitivity $f'_i$ and buckets of different
sizes $S_j$ introduces the new challenge of finding the best bucket
setting that can fulfil the requirement $f'_i$ for all $x_i$'s. Even
for a given bucket setting, it is non-trivial to validate whether
there is a valid record assignment to the buckets. Secondly, the
number of feasible bucket settings of the form $\langle
(S_1,b_1),\cdots,(S_q,b_q)\rangle$ is huge, rendering it prohibitive
to enumerate all bucket settings. For example, suppose that $S_1$
and $S_2$ are chosen from the range of $[3,20]$, and
$|T|=1,000,000$, there are a total of 2,077,869 feasible bucket
settings of the form $(S_1,b_1)$ and $(S_2,b_2)$. This number will
be much larger if $q>2$. Finally, the number of distinct bucket
sizes $q$ is unknown in advance and must be searched.\end{remark}

Section 4 presents an algorithm for validating a two-size bucket
setting. Section 5 presents an efficient algorithm for the optimal
two-size bucket setting problem with \emph{guaranteed} optimality,
and a heuristic algorithm for the multi-size bucket setting problem.

\section{Validating Two-Size Bucket Setting}\label{two-size}
Let $Valid(B,T,F')$ denote a function that tests if a bucket setting
$B$ is valid. We assume that the number of occurrence $o_i$ for
$x_i$ in $T$ has been collected, $1\leq i\leq m$. In Section 4.1, we
consider buckets having the same size and we give an $O(m)$ time and
space algorithm for evaluating $Valid(B,T,F')$. In Section 4.2, we
consider buckets having two different sizes and give an $O(m)$ time
and space algorithm for $Valid(B,T,F')$. In both cases, we give a
linear time algorithm for finding a valid record assignment for a
valid bucket setting.

\subsection{One-Size Bucket Setting}

Let $B=\{g_0,\cdots,g_{b-1}\}$ be a set of $b$ buckets of the
\emph{same} size $S$. To validate this bucket setting, we introduce
a round-robin assignment of records to buckets.

\emph{\textbf{Round-Robin Assignment (RRA)}}: For each value $x_i$,
$1\leq i\leq m$, we assign the $t$-th record of $x_i$ to the bucket
$g_s$, where $s=(o_1+\cdots+o_{i-1} + t) \mod b$, where $o_i$ is the
number of occurrence of $x_i$ in $T$. In other words, the records
for $x_i$ are assigned to the buckets in a round-robin manner; the
order in which $x_i$ is considered by RRA is not important. It is
easy to see that the number of records for $x_i$ assigned to a
bucket is either $\lfloor |o_i|/b \rfloor$ or $\lceil |o_i|/b
\rceil$. The next lemma gives a sufficient and necessary condition
for $Valid(B,T,F')=true$.

\begin{lemma}[Validating one-size bucket setting]\label{l1}
Let $B$ be a set of $b$ buckets of size $S$ such that $|T|=s(B)$.
The following are equivalent: (1) $Valid(B,T,F')=true$. (2) There is
a valid RRA from $T$ to $B$ wrt $F'$. (3) For each $SA$ value $x_i$,
$\frac{\lceil o_i/b \rceil}{S} \leq f'_i $. (4) For each $SA$ value
$x_i$, $o_i\leq \lfloor f'_iS \rfloor b$.
\end{lemma}

\begin{proof}
We show $4 \Rightarrow 3 \Rightarrow 2 \Rightarrow 1 \Rightarrow 4$.
Observe that if $r$ is a real number and $i$ is an integer, $r\le i$
if and only if $\lceil r \rceil \le i$, and $i\le r$ if and only if
$i \le \lfloor r \rfloor$. Then the following rewriting holds.


$\frac{\lceil o_i/b \rceil}{S} \leq f'_i \Leftrightarrow \lceil
o_i/b \rceil \leq f'_iS \Leftrightarrow \lceil o_i/b \rceil \leq
\lfloor f'_iS \rfloor \Leftrightarrow o_i/b \leq \lfloor f'_iS
\rfloor \Leftrightarrow o_i \leq \lfloor f'_iS \rfloor b$. This
shows the equivalence of 4 and 3.

To see $3 \Rightarrow 2$, observe that $\frac{\lceil o_i/b
\rceil}{S}$ is the maximum frequency of $x_i$ in a bucket generated
by RRA. Condition 3 implies that this assignment is valid. $2
\Rightarrow 1$ follows because every valid RRA is a valid
assignment. To see $1 \Rightarrow 4$, observe that $F'$-privacy
implies that the number of occurrence of $x_i$ in a bucket of size
$S$ is at most $\lfloor f'_iS \rfloor$. Thus for any valid
assignment, the total number of occurrence $o_i$ in the $b$ buckets
of size $S$ is no more than $\lfloor f'_iS \rfloor b$.
\end{proof}

\subsection{Two-Size Bucket Setting}\label{two-size-validating}
Now we consider a two-size bucket setting of the form $\langle
B_1(S_1,b_1),B_2(S_2,b_2)\rangle$.
The next lemma follows trivially.


\begin{lemma} \label{l2} $Valid(B_1 \cup B_2,T,F')=true$
if and only if there is a partition of $T$, $\{T_1,T_2\}$, such that
$Valid(B_1,T_1,F')=true$ and $Valid(B_2,T_2,F')=true$.
\end{lemma}

\begin{definition}\label{d1}
Given $F'$-privacy, for each $x_i$ and for $j=1,2$, we define
$u_{ij} = \lfloor f'_i S_j \rfloor b_j$ and $a_{ij}=min\{u_{ij},
o_i\}$.
\end{definition}

From Lemma \ref{l1}(4), $u_{ij}$ is the upper bound on the number of
records for $x_i$ that can be allocated to $B_j$ without violating
$f'_i$-privacy, assuming unlimited supply of $x_i$ records. $a_{ij}$
is the upper bound, assuming the \emph{actual} supply of $x_i$
records, i.e., $o_i$. The next theorem gives the condition for
$Valid(B_1 \cup B_2,T,F')=true$.

\begin{theorem}[Validating two-size bucket setting]\label{th1}
$Valid(B_1 \cup B_2,T,F')=true$ if and only if all of the following
conditions hold:
\begin{eqnarray}
   \forall i : a_{i1} + a_{i2} \ge o_i & (Privacy\ Constraint (PC)) \label{a3}\\
   j=1,2 : \sum_i a_{ij} \geq s(B_j) & (Fill\ Constraint (FC))) \label{a4}\\
        |T|=s(B_1)+s(B_2) & (Capacity \ Constraint (CC))) \label{a5}
 \end{eqnarray}
\end{theorem}

\begin{proof}
Intuitively, Equation (\ref{a3}) says that the number of occurrence
of $x_i$ does not exceed the upper bound $a_{i1}+a_{i2}$ imposed by
$F'$-privacy on all buckets collectively, thus, the name Privacy
Constraint. Equation (\ref{a4}) says that under this upper bound
constraint it is possible to fill up the buckets in $B_j$ without
leaving unused slots, thus, the name Fill Constraint. Equation
(\ref{a5}) says that the total bucket capacity matches the data
cardinality, thus the name Capacity Constraint. Clearly, all these
conditions are necessary for a valid assignment. The sufficiency
proof is given by the algorithm in the next subsection that finds a
valid assignment of the records in $T$ to the buckets in $B_1$ and
$B_2$, assuming that the above conditions hold.
\end{proof}

In the rest of the paper, PC, FC, and CC denote Privacy Constraint,
Fill Constraint, and Capacity Constraint in Theorem \ref{th1}.


\begin{corollary}
For a set buckets $B$ with at most two bucket sizes,
$Valid(B,T,F')=true$ can be tested in $O(m)$ time and $O(m)$ space.
\end{corollary}

\subsection{Record Partitioning}\label{assignment}
Suppose that PC, FC and CC in Theorem \ref{th1} hold. We show how to
find a partition $\{T_1,T_2\}$ of $T$ such that
$Valid(B_1,T_1,F')=true$ and $Valid(B_2,T_2,F')=true$. This provides
the sufficiency proof for Theorem \ref{th1} because Lemma \ref{l2}
implies $Valid(B_1\cup B_2,T,F')=true$. By finding the partition
$\{T_1, T_2\}$, we also provide an algorithm for assigning records
from $T$ to the buckets in $B_1 \cup B_2$, that is, simply applying
RRA to each of $(T_j,B_j)$, $j=1,2$.


The partition $\{T_1,T_2\}$ can be created as follows. For each $SA$
value $x_i$, initially $T_1$ contains any $a_{i1}$ records and $T_2$
contains the remaining $o_i -a_{i1}$ records for $x_i$. Since
$a_{i1}\leq u_{i1}$, Lemma \ref{l1}(4) holds on $(T_1,B_1)$. (Note
that in this case, $o_i$ in Lemma \ref{l1} is the number of
occurrence of $x_i$ in $T_1$.) PC implies that the number of
occurrence of $x_i$ in $T_2$, i.e., $o_i -a_{i1}$, is no more than
$a_{i2}$, therefore, Lemma \ref{l1}(4) also holds on $(T_2,B_2)$. FC
implies $|T_1|\geq s(B_1)$. If $|T_1|=s(B_1)$, $|T_2|=s(B_2)$ (i.e.,
CC), from the above discussion and Lemma \ref{l1},
$Valid(B_1,T_1,F')=true$ and $Valid(B_2,T_2,F')=true$. We are done.

We now assume $|T_1| > s(B_1)$, thus $|T_2| < s(B_2)$. We need to
move $|T_1|- s(B_1)$ records from $T_1$ to $T_2$ without exceeding
the upper bound $a_{i2}$ for $T_2$. FC implies that such moves are
possible because there must be some $x_i$ for which less than
$a_{i2}$ records are found in $T_2$. For such $x_i$, we move records
of $x_i$ from $T_1$ to $T_2$ until the number of records for $x_i$
in $T_2$ reaches $a_{i2}$ or until $|T_2| = s(B_2)$, whichever comes
first. Since we move a record for $x_i$ to $T_2$ only when there are
less than $a_{i2}$ records for $x_i$ in $T_2$, the condition of
Lemma \ref{l1}(4) is preserved  on $(T_2,B_2)$. Clearly, moving a
record out of $T_1$ always preserves the condition of Lemma
\ref{l1}(4) on $(T_1,B_1)$. As long as $|T_2| < s(B_2)$, the above
argument can be repeated to move more records from $T_1$ to $T_2$.

Eventually, we have $|T_2| = s(B_2)$, so $Valid(B_1,T_1,F')=true$
and $Valid(B_2,T_2,F')=true$. The $\{T_1,T_2\}$ is the partition
required.

\begin{figure}[h]
\subfigure[The bucket for $B_1$]{
\begin{minipage}[t]{1\linewidth}
\centering
\begin{tabular}{|c|c|c|c|c|c|c|c|c|c|c|c|c|c|c|}
  \hline
 $1$ & $2$ & $3$ & $4$ & $5$ & $6$ & $7$ & $8$ & $9$ & $10$ & $11$ & $12$ & $13$ & $14$\\ \hline
 \end{tabular}
\end{minipage}}
\centering
 \subfigure[The buckets for $B_2$]{
\begin{minipage}[t]{0.45\linewidth}
\centering
\begin{tabular}{|c|c|c|c|c|}\hline
 $9$ & $10$ & $12$ & $13$\\ \hline
 $9$ & $11$ & $12$ & $14$ \\ \hline
 $9$ & $11$ & $13$ & $14$ \\ \hline
 $9$ & $11$ & $13$ & $14$ \\ \hline
 $9$ & $11$ & $13$ & $14$ \\ \hline
 $10$ & $11$ & $13$ & $14$\\ \hline
 $10$ & $12$ & $13$ & $14$\\ \hline
 $10$ & $12$ & $13$ & $14$\\ \hline
 $10$ & $12$ & $13$ & $14$\\ \hline
   \end{tabular}
\end{minipage}}

\caption{The record assignment for Example
\ref{ex:F'}}\label{tab:example2}
\end{figure}

\begin{example} \label{ex:F'} Suppose $f'_i= 2\times f_i+0.05$. Consider a table $T$ containing 50 records
with $o_i$ for $x_i$ as follows:
 \btab \ppos
\> \>  $x_1$-$x_8$: $o_i=1$, $f_i = 0.02$ and $f'_i = 0.09$. \\
\> \> $x_9$-$x_{12}$: $o_i=6$, $f_i = 0.12$ and $f'_i = 0.29$. \\
\> \>  $x_{13}$-$x_{14}$: $o_i=9$, $f_i = 0.18$ and $f'_i = 0.41$.
 \etab
Consider the bucket setting $B_1(S_1=4,b_1=9),B_2(S_2=14,b_2=1)$.
Note CC in Theorem \ref{th1} holds. Let us compute $a_{i1}$ and
$a_{i2}$.

$a_{i1}=min\{u_{i1},o_i\}$: For $x_1$-$x_8$, $u_{i1}=\lfloor f'_i
S_1 \rfloor b_1=\lfloor 0.09 \times 4 \rfloor \times 9= 0$, so
$a_{i1}=0$. For $x_9$-$x_{12}$, $u_{i1}=\lfloor 0.29\times 4 \rfloor
\times 9=9$, \textbf{$a_{i1}=6$.} For $x_{13}$-$x_{14}$,
$u_{i1}=\lfloor 0.41 \times 4 \rfloor \times 9=9$, $a_{i1}=9$.

$a_{i2}=min \{u_{i2},o_i\}$: For $x_1$-$x_8$, $u_{i2}=\lfloor f'_i
S_2 \rfloor b_2=\lfloor 0.09 \times 14 \rfloor \times 1= 1$,
$a_{i2}=1$. For $x_9$-$x_{12}$, $u_{i2}=\lfloor 0.29\times 14
\rfloor \times 1=4$, $a_{i2}=4$. For $x_{13}$-$x_{14}$,
$u_{i2}=\lfloor 0.41 \times 14 \rfloor \times 1= 5$, $a_{i2}=5$.

It can be verified that PC and FC in Theorem \ref{th1} hold. To find
the partitioning $\{T_1,T_2\}$, initially $T_1$ contains $a_{i1}=0$
record for each of $x_1$-$x_8$, $a_{i1}=6$ records for each of
$x_9$-$x_{12}$, and $a_{i1}=9$ records for each of
$x_{13}$-$x_{14}$. $T_2$ contains the remaining records in $T$.
Since $T_1$ contains 42 records, but $s(B_1)=36$, we need to move 6
records from $T_1$ to $T_2$ without exceeding the upper bound
$a_{i2}$ for $T_2$. This can be done by moving one record for each
of $x_9-x_{14}$ from $T_1$ to $T_2$. Figure \ref{tab:example2} shows
a record assignment generated by RRA for $(B_1,T_1)$ and
$(B_2,T_2)$.
\end{example}

\section{Finding Optimal Bucket Settings}\label{optimal}

We now present an efficient algorithm for finding the optimal bucket
setting. Section 5.1 presents an exact solution for the two-size
bucket setting problem. Section 5.2 presents a heuristic solution
for the multi-size bucket setting problem.

\subsection{Algorithms for Two-Size Bucket Settings}
Given $T$ and $F'$-privacy, we want to find the valid bucket setting
of the form $\langle B_1(S_1,b_1), B_2(S_2,b_2)\rangle$, where
$b_j\geq 0$ and $S_1<S_2$, such that the following loss is minimized
\begin{equation} Loss(B_1\cup
B_2)=b_1 (S_1-1)^2 + b_2 (S_2-1)^2 \label{lossforB1B2}
\end{equation}
One approach is applying Theorem \ref{th1} to validate each feasible
bucket setting $(B_1,B_2)$, but this is inefficient because the
number of such bucket settings can be huge (Remark \ref{remark2}).
We present a more efficient algorithm that prunes the bucket
settings that are not valid or do not have the minimum loss. Observe
that $f'_i$-privacy implies that a record for $x_i$ must be placed
in a bucket of size at least $\lceil1/f'_i \rceil$; therefore, the
minimum size for $S_1$ and $S_2$ is $M=min_i\{\lceil 1/f'_i
\rceil\}$. The maximum bucket size $M'$ for $S_1$ and $S_2$ is
constrained by the maximum loss allowed. We assume that $M'$ is
given, where $M' > M$. We consider only $(S_1,S_2)$ such that $M\leq
S_1 < S_2\leq M'$. Note that a valid bucket setting may not exist in
this range of size.

\subsubsection{Indexing Bucket Settings}
We first present an ``indexing" structure for feasible bucket
settings to allow a direct access to any feasible bucket setting. We
say that a pair $(b_1,b_2)$ is \emph{feasible} (resp. \emph{valid})
wrt $(S_1,S_2)$ if the bucket setting $\langle
B_1(S_1,b_1),B_2(S_2,b_2)\rangle$ is feasible (resp. valid). A valid
pair $(b_1,b_2)$ is \emph{optimal} wrt $(S_1,S_2)$ if $Loss(B_1\cup
B_2)$ is minimum among all valid pairs wrt $(S_1,S_2)$. We define
$\Gamma(S_1,S_2)$ to be the list of all feasible $(b_1,b_2)$ in the
descending order of $b_1$, thus, in the ascending order of $b_2$.
Intuitively, an earlier bucket setting has more smaller buckets,
thus, a smaller $Loss$, than a later bucket setting. Below, we show
that the $i$-th pair in $\Gamma(S_1,S_2)$ can be generated
\emph{directly} using the position $i$ without scanning the list. We
will use this property to locate all valid pairs by a binary search
on $\Gamma(S_1,S_2)$ without storing the list. To this end, it
suffices to identify the first and last pairs in $\Gamma(S_1,S_2)$,
and the increments of $b_1$ and $b_2$ between two consecutive pairs.

The first pair in $\Gamma(S_1,S_2)$, denoted $(b_1^0,b_2^0)$, has
the largest possible $b_1$ such that $S_1 b_1 + S_2 b_2 =|T|$. So
$(b_1^0,b_2^0)$ is the solution to the following integer linear
program:
\begin{equation}
min \{b_2 \mid S_1 b_1 + S_2 b_2 =|T|\}\label{b0}
\end{equation}
$b_1$ and $b_2$ are variables of non-negative integers and
$S_1,S_2,|T|$ are constants.

Next, consider two consecutive pairs $(b_1,b_2)$ and $(b_1
-\Delta_1,b_2+\Delta_2)$ in $\Gamma(S_1,S_2)$. Since $S_1 b_1 + S_2
b_2 =|T|$ and $S_1(b_1 -\Delta_1) + S_2(b_2+\Delta_2)=|T|$, $S_1
\Delta_1=S_2\Delta_2$. Since $\Delta_1$ and $\Delta_2$ are the
smallest positive integers such that this equality holds, $S_2
\Delta_2$ must be the least common multiple of $S_1$ and $S_2$,
denoted by $LCM(S_1,S_2)$.
$\Delta_2$ and $\Delta_1$ are then given by
\begin{equation}
\Delta_2 = LCM(S_1,S_2)/S_2, \ \Delta_1 = LCM(S_1,S_2)/S_1
\label{delta}
\end{equation}
%
Therefore, the $i$th pair in $\Gamma(S_1,S_2)$ has the form
$(b_1^0-i*\Delta_1,b_2^0+i*\Delta_2)$, where $i\geq 0$. The last
pair has the maximum $i$ such that $0\leq b_1^0 -i*\Delta_1 <
\Delta_1$, or $b_1^0 /\Delta_1 -1 < i \leq b_1^0/\Delta_1$. The only
integer $i$ satisfying this condition is given by
\begin{eqnarray}
k=\lfloor b_1^0/\Delta_1 \rfloor \label{k}
 \end{eqnarray}

\begin{lemma} \label{gammalist}
$\Gamma(S_1,S_2)$ has the form
\begin{equation}
(b_1^0,b_2^0), (b_1^0-\Delta_1,b_2^0+\Delta_2), \cdots,
(b_1^0-k*\Delta_1,b_2^0+k*\Delta_2)
\end{equation}
where $b_1^0,b_2^0, \Delta_1,\Delta_2, k$ are defined in Equations
(\ref{b0}-\ref{k}).
\end{lemma}

\begin{remark} \label{remark3}
$\Gamma(S_1,S_2)$ in Lemma \ref{gammalist} has several important
properties for dealing with a large data set. Firstly, we can access
the $i$-th element of $\Gamma(S_1,S_2)$ without storing or scanning
the list. Secondly, we can represent any sublist of
$\Gamma(S_1,S_2)$ by a bounding interval $[i,j]$ where $i$ is the
starting position and $j$ is the ending position of the sublist.
Thirdly, the common sublist of two sublists $L$ and $L'$ of
$\Gamma(S_1,S_2)$, denoted by $L\cap L'$, is given by the
intersection of the bounding intervals of $L$ and $L'$.
\end{remark}

\begin{example} \label{ex:FL} Let $|T|=28, S_1=2,S_2=4$. $LCM(S_1,S_2)=4$.
$\Delta_2=4/4=1$ and $\Delta_1 =4/2=2$. $b_1^0=14, b_2^0=0$.
$k=\lfloor 14/2 \rfloor =7$. $\Gamma(S_1,S_2)$ is (14,0), (12,1),
(10,2), (8,3), (6,4), (4,5), (2,6), (0,7).
\end{example}


The length $k$ of $\Gamma(S_1,S_2)$, given by Equation (\ref{k}), is
proportional to the cardinality $|T|$. $b_1^0$ is as large as
$|T|/S_1$ (when $b_2^0=0$) and $\Delta_1$ is no more than $S_2$.
Thus $k$ is as large as $|T|/(S_1S_2)$. With $S_1$ and $S_2$ being
small, $k$ is proportional to $|T|$. Therefore, examining all pairs
in $\Gamma(S_1,S_2)$ is not scalable.
In the rest of this section, we explore two pruning strategies to
prune unpromising pairs $(b_1,b_2)$ in $\Gamma(S_1,S_2)$, one based
on loss minimization and one based on privacy requirement.

\subsubsection{Loss-Based Pruning} Our first strategy is pruning the
pairs in $\Gamma(S_1,S_2)$ that do not have the minimum loss wrt
$(S_1,S_2)$, by exploiting the following monotonicity of $Loss$,
which follows from the descending order of $b_1$, $S_1 < S_2$, and
Equation (\ref{lossforB1B2}).

\begin{lemma}[Monotonicity of loss]\label{l3}
If $(b_1,b_2)$ precedes $(b'_1,b'_2)$ in $\Gamma(S_1,S_2)$.
$Loss(B_1\cup B_2) < Loss(B'_1\cup B'_2)$, where $B_j$ contains
$b_j$ buckets of size $S_j$, and $B'_j$ contains $b'_j$ buckets of
size $S_j$, $j=1,2$.
\end{lemma}

Thus the first valid pair in $\Gamma(S_1,S_2)$ is the optimal pair
wrt $(S_1,S_2)$. Lemma \ref{l3} can also be exploited to prune pairs
across different $(S_1,S_2)$. Let $Best_{loss}$ be the minimum loss
found so far and $(S_1,S_2)$ be the next pair of sizes to be
considered. From Lemma \ref{l3}, all the pairs in $\Gamma(S_1,S_2)$
that have a loss less than $Best_{loss}$ must form a prefix of
$\Gamma(S_1,S_2)$. Let $(b_1^*,b_2^*)$ be the cutoff point of this
prefix, where $b_1^*=b_1^0-k^*
* \Delta_1$ and $b_2^*=b_2^0+k^*
* \Delta_2$. $k^*$ is the maximum integer satisfying $b_1^*(S_1-1)^2 + b_2^*(S_2-1)^2 < Best_{loss}$.
$k^*$ is given by
\begin{equation}
k^*=max\{0,\lfloor \frac{Best_{loss} - b_1^0(S_1 -1)^2 -b_2^0 (S_2
-1)^2}{\Delta_2 (S_2 -1)^2 -\Delta_1 (S_1 -1)^2} \rfloor \}
\label{t}
\end{equation}

The next lemma revises $\Gamma(S_1,S_2)$ by the cutoff point based
on $Best_{loss}$.

\begin{lemma}[Loss-based pruning]
Let $Best_{loss}$ be the minimum loss found so far and let
$(S_1,S_2)$ be the next pair of sizes to consider. Let
$k'=min\{k,k^*\}$, where $k$ is given by Equation (\ref{k}) and
$k^*$ is given by Equation (\ref{t}). Let $\Gamma'(S_1,S_2)$ denote
the prefix of $\Gamma(S_1,S_2)$ that contains the first $k'+1$
pairs. It suffices to consider $\Gamma'(S_1,S_2)$. \label{l44}
\end{lemma}

In the rest of this section, $\Gamma'$ denotes $\Gamma'(S_1,S_2)$
when $S_1$ and $S_2$ are clear from context.

\subsubsection{Privacy-Based Pruning}
From Lemma \ref{l3}, the optimal pair wrt $(S_1,S_2)$ is the first
valid pair in $\Gamma'$. Our second strategy is to locate the first
valid pair in $\Gamma'$ \emph{directly} by exploiting a certain
monotonicity property of the condition for a valid pair. First, we
introduce some terminology. Consider any sublist $L$ of $\Gamma'$
and any boolean condition $C$ on a pair. $H(C,L)$ denotes the set of
all pairs in $L$ on which $C$ holds, and $F(C,L)$ denotes the set of
all pairs in $L$ on which $C$ fails.  $C$ is \emph{monotone} in $L$
if whenever $C$ holds on a pair in $L$, it holds on all later pairs
in $L$, and \emph{anti-monotone} in $L$ if whenever $C$ fails on a
pair in $L$, it fails on all later pairs in $L$. A monotone $C$
splits $L$ into two sublists $F(C,L)$ and $H(C,L)$ in that order,
and an anti-monotone $C$ splits $L$ into two sublists $H(C,L)$ and
$F(C,L)$ in that order. Therefore, if we can show that FC and PC in
Theorem \ref{th1} are monotone or anti-monotone, we can locate all
valid pairs in $\Gamma'$, i.e., those satisfying both FC and PC, by
a binary search over $\Gamma'$. We consider FC and PC separately.

\medskip

 \textbf{Monotonicity of FC}. Let $FC(S_1)$
denote FC for $j=1$, and $FC(S_2)$ denote FC for $j=2$. Note that
$H(FC,\Gamma')$ is given by $H(FC(S_1),\Gamma')\cap
H(FC(S_2),\Gamma')$.

\begin{lemma}[Monotonicity of FC]\label{l4}
$FC(S_1)$ is monotone in $\Gamma'$ and $FC(S_2)$ is anti-monotone in
$\Gamma'$.
\end{lemma}

\begin{proof} We rewrite FC as
\begin{eqnarray}
\sum_i min_i \{\lfloor f'_iS_1 \rfloor b_1,o_i\} \geq S_1 b_1
\label{a6}\\
 \sum_i min_i \{\lfloor f'_iS_2 \rfloor b_2,o_i\} \geq
S_2 b_2 \label{a7}
\end{eqnarray}
Assume that $(b_1,b_2)$ precedes $(b'_1,b'_2)$ in $\Gamma'$. Then
$b_1 > b'_1$ and $b_2 < b'_2$. As $b_1$ decreases to $b'_1$, both
$\lfloor f'_iS_1 \rfloor b_1$ and $S_1 b_1$ decreases by a factor by
$b'_1/b_1$, but $o_i$ remains unchanged. Therefore, if Equation
(\ref{a6}) holds for $(b_1,b_2)$, it holds for $(b'_1,b'_2)$ as
well; so Equation (\ref{a6}) is monotone on $\Gamma'$. For a similar
reason, if Equation (\ref{a7}) fails on $(b_1,b_2)$, it remains to
fail on $(b'_1,b'_2)$ as well; thus Equation (\ref{a7}) is
anti-monotone on $\Gamma'$.
\end{proof}

\textbf{Monotonicity of PC}. Let $PC(x_i)$ denote PC for $x_i$.
$H(PC,\Gamma')$ is given by $\cap_i H(PC(x_i),\Gamma')$. To compute
$H(PC(x_i), \Gamma')$, we rewrite $PC(x_i)$ as
\begin{equation}
 min\{\lfloor f'_iS_1 \rfloor b_1,o_i\} + min\{\lfloor f'_iS_2
\rfloor b_2,o_i\} \geq o_i \label{a8}
\end{equation}
Since $b_1$ is decreasing and $b_2$ is increasing in $\Gamma'$,
$\lfloor f'_iS_1 \rfloor b_1\geq o_i$ is anti-monotone and $\lfloor
f'_iS_2 \rfloor b_2 \geq o_i$ is monotone in $\Gamma'$. Note
Equation (\ref{a8}) holds in $H(\lfloor f'_iS_1 \rfloor b_1\geq o_i,
\Gamma')$ and $H(\lfloor f'_iS_2 \rfloor b_2 \geq o_i, \Gamma')$.
%
%

Let us consider the remaining part of $\Gamma'$, denoted by
$\Gamma'(x_i)$:
\[
F(\lfloor f'_iS_1 \rfloor b_1 \geq o_i,\Gamma') \cap F(\lfloor
f'_iS_2 \rfloor b_2 \geq o_i, \Gamma'). \] In this part, Equation
(\ref{a8}), thus $PC(x_i)$, degenerates into
\begin{equation}
\lfloor f'_iS_1 \rfloor b_1 + \lfloor f'_iS_2 \rfloor b_2 \geq o_i
\label{a9}
\end{equation}
Consider
\begin{equation}
\lfloor f'_iS_2 \rfloor \Delta_2 \geq \lfloor f'_iS_1 \rfloor
\Delta_1 \label{a10}
\end{equation}
and any two consecutive pairs $(b_1,b_2)$ and
$(b_1-\Delta_1,b_2+\Delta_2)$ in $\Gamma'(x_i)$. If Equation
(\ref{a10}) holds, Equation (\ref{a9}) holding on $(b_1,b_2)$
implies that it holds on $(b_1-\Delta_1,b_2+\Delta_2)$, thus,
Equation (\ref{a9}) is monotone; if Equation (\ref{a10}) fails,
Equation (\ref{a9}) failing on $(b_1,b_2)$ implies that it fails on
$(b_1-\Delta_1,b_2+\Delta_2)$, thus, Equation (\ref{a9}) is
anti-monotone. Recall that in $\Gamma'(x_i)$, $PC(x_i)$ degenerates
into Equation (\ref{a9}). The next lemma summarizes the above
discussion.

\begin{lemma}[Monotonicity of PC]\label{l7}
(i) $\lfloor f'_iS_1 \rfloor b_1\geq o_i$ is anti-monotone in
$\Gamma'$ and $\lfloor f'_iS_2 \rfloor b_2 \geq o_i$ is monotone in
$\Gamma'$. (ii) If Equation (\ref{a10}) holds, $PC(x_i)$ is monotone
in $\Gamma'(x_i)$, and if Equation (\ref{a10}) fails, $PC(x_i)$ is
anti-monotone in $\Gamma'(x_i)$.
\end{lemma}


\begin{corollary}\label{c3}
$H(PC(x_i),\Gamma')$ consists of $H(\lfloor f'_iS_1 \rfloor b_1\geq
o_i, \Gamma')$, $H(PC(x_i),\Gamma'(x_i))$, and $H(\lfloor f'_iS_2
\rfloor b_2 \geq o_i,\Gamma')$.
\end{corollary}

\subsubsection{Algorithms}
The next theorem gives a computation of all pairs in  $\Gamma'$
satisfying both PC and FC, i.e., all valid pairs in $\Gamma'$.

\begin{theorem}[Computing all valid pairs in $\Gamma'$]\label{th3}
Let $\Gamma^*$ be the intersection of $H(FC(S_1),\Gamma')$,
$H(FC(S_2),\Gamma')$, and $\cap_i H(PC(x_i),\Gamma')$. (i)
$\Gamma^*$ contains exactly the valid pairs in $\Gamma'$. (ii) The
first pair in $\Gamma^*$ (if any) is the optimal pair wrt
$(S_1,S_2)$. (iii) $\Gamma^*$ can be computed in $O(m \log |T|)$
time and $O(m)$ space.
\end{theorem}

\begin{proof} (i) follows from Theorem \ref{th1}.
From Lemma \ref{l3}, the first pair in $\Gamma^*$ has the minimum
loss wrt $(S_1,S_2)$. To see (iii), the monotonicities in Lemma
\ref{l4} and Lemma \ref{l7}, and Corollary \ref{c3}, imply that each
sublist involved in computing $\Gamma^*$ can be found by a binary
search over $\Gamma'$, which takes $O(m \log |T|)$ time (note that
the length $k'$ of $\Gamma'$ is no more than $|T|$). Note that
intersecting two sublists takes $O(1)$. The $O(m)$ space follows
from the fact that each sublist is represented by its bounding
interval and any element of $\Gamma'$ examined by a binary search
can be generated based on its position without storing the list.
\end{proof}

\begin{algorithm}[tb]
\caption{Optimal Two-Size Bucketing} \label{alg:2-size2}
\textbf{TwoSizeBucketing$(T,F',M,M')$}\\
Input: $T$, $1\leq i\leq m$, $F'$, $M,M'$\\
Output: the optimal bucket setting $\langle
(S_1,b_1),(S_2,b_2)\rangle$
\begin{algorithmic} [1]
 \STATE compute $o_i$, $1\leq i\leq m$
 \STATE $Best_{loss} \leftarrow \infty$
 \STATE $Best_{setting} \leftarrow NULL$
 \FORALL {$\{S_1=M$; $S_1\leq M'-1$; $S_1++\}$}
    \FORALL {$\{S_2=S_1+1$; $S_2 \leq M'$; $S_2++\}$}
      \STATE compute $\Gamma^*$ using Theorem \ref{th3}
            \IF {$\Gamma^*$ is not empty}
         \STATE let $(b_1,b_2)$ be the first pair in $\Gamma^*$
         \STATE let $B_j$ be the set of $b_j$ buckets of size $S_j$, $j=1,2$
         \IF {$Best_{loss} > Loss(B_1 \cup B_2)$}
             \STATE $Best_{setting} \leftarrow \langle (S_1,b_1),(S_2,b_2)\rangle$
             \STATE $Best_{loss} \leftarrow Loss(B_1 \cup B_2)$
         \ENDIF
      \ENDIF
    \ENDFOR
 \ENDFOR
 \STATE return $Best_{setting}$
\end{algorithmic}
\end{algorithm}

Algorithm \ref{alg:2-size2} presents the algorithm for finding the
optimal two-size bucket setting based on Theorem \ref{th3},
\emph{TwoSizeBucketing}. The input consists of a table $T$, a
privacy parameter $F'$, and the minimum and maximum bucket sizes $M$
and $M'$. Line 1 computes $o_i$ in one scan of $T$. Lines 2 and 3
initialize $Best_{loss}$ and $Best_{setting}$. Lines 4 and 5 iterate
through all pairs $(S_1,S_2)$ with $M\leq S_1 < S_2\leq M'$. For
each pair $(S_1,S_2)$, Line 6 computes the list $\Gamma^*$ using
Theorem \ref{th3}. Lines 8-12 compute $Loss$ of the first pair in
$\Gamma^*$ and update $Best_{loss}$ and $Best_{setting}$ if
necessary. Line 13 returns $Best_{setting}$. The algorithm uses both
loss-based pruning and privacy-based pruning. The former is through
the prefix $\Gamma'$ obtained by the upper bound $Best_{loss}$ as
computed in Lemma \ref{l44}, and the latter is through the binary
search of valid pairs implicit in the computation of $\Gamma^*$. To
tighten up $Best_{loss}$, Lines 4 and 5 examine smaller sizes
$(S_1,S_2)$ before larger ones.

\subsection{Algorithms for Multi-Size Bucket Settings}
A natural next step is to extend the solution for the two-size
problem to the multi-size problem. To do so, we must extend Theorem
\ref{th1} to validate a three-size bucket setting. The next example
shows that this does not work.

\begin{example}\label{ex:3-size} Let $|B_1| = |B_2| = 20$, $|B_3| = 30$, and
$|T|=70$. There are 11 values $x_1,\cdots,x_{11}$: $o_i=5$ for
$1\leq i \leq 10$, and $o_{11}=20$. Suppose that for $1\leq i\leq
10$, $a_{i1}=a_{i2}=0$, $a_{i3}=5$, and
$a_{11,1}=a_{11,2}=a_{11,3}=20$. The following extended version of
PC, FC and CC in Theorem \ref{th1}: $\forall i : a_{i1} + a_{i2} +
a_{i3} \ge o_i$; for $j=1,2,3$, $\sum_i a_{ij} \geq |B_j|$;
$|T|=|B_1|+|B_2|+|B_3|$. However, there is no valid record
assignment to these buckets. Note that, for $1\leq i\leq 10$,
$a_{i1}=a_{i2}=0$, none of the records for $x_i$ can be assigned to
the buckets for $B_1$ or $B_2$. So the 50 records for $x_i$, $1\leq
i\leq 10$, must be assigned to the buckets for $B_3$, but $B_3$ has
a capacity of 30.
\end{example}

Our solution is recursively applying \emph{TwoSizeBucketing} to
reduce $Loss$. This algorithm, \emph{MultiSizeBucketing}, is given
in Algorithm \ref{alg:heuristic}. The input consists of $T$, a set
of records, $B$, a set of buckets of the same size, and $F',M,M'$ as
usual, where $|T|=s(B)$. The algorithm applies
\emph{TwoSizeBucketing} to find the optimal two-size bucket setting
$(B_1,B_2)$ for $T$ (Line 1). If $Loss(B_1 \cup B_2) < Loss(B)$,
Line 3 partitions the records of $T$ into $T_1$ and $T_2$ between
$B_1$ and $B_2$.  $RecordPartition(T,B_1,B_2)$ is the record
partition procedure discussed in Section \ref{assignment}. Lines 4
and 5 recur on each of $(T_1,B_1)$ and $(T_2,B_2)$. If $Loss(B_1
\cup B_2) \geq Loss(B)$, Line 7 returns the current bucket setting
$B$ for $T$.

\begin{algorithm}[tb]
\caption{Heuristic Multi-Size Bucketing} \label{alg:heuristic}
\textbf{MultiSizeBucketing}$(T,B,F',M,M')$\\
Input: $T$, $B$, $F',M,M'$\\
Output: a bucket setting $\langle B_1,\cdots,B_q\rangle$ and
$T_1,\cdots,T_q$, where $T_j$ is a set of records for $B_j$, $1\leq
j\leq q$
\begin{algorithmic} [1]
     \STATE $\langle B_1,B_2\rangle \leftarrow TwoSizeBucketing(T,F',M,M')$
      \IF {$Loss(B_1\cup B_2)<Loss(B)$}
       \STATE $(T_1,T_2) \leftarrow RecordPartition(T,B_1,B_2)$ (Section
     \ref{assignment})
        \STATE $MultiSizeBucketing(T_1,B_1,F',M,M')$
        \STATE $MultiSizeBucketing(T_2,B_2,F',M,M')$
     \ELSE
        \STATE return($T,B$)
     \ENDIF
\end{algorithmic}
\end{algorithm}

\section{Additional Auxiliary Information}\label{extension}
Dealing with an adversary armed with additional auxiliary
information is one of the hardest problems in data privacy. As
pointed out by \cite{Kifer2011}, there is no free lunch in data
privacy. Thus, instead of dealing with all types of auxiliary
information, we consider two previously identified attacks, namely,
corruption attack \cite{TaoCorr08} and negative association attack
\cite{Kifer:2009:APD:1559845.1559861}\cite{Li09}. To focus on the
main idea, we consider $F'$-privacy such that $f'_i$ is the same for
all $x_i$'s. In this case, $F'$-privacy degenerates into
$\ell$-diversity with $\ell=\lceil 1/f'_i\rceil$ and the solution in
Section 5.1 returns buckets of size $S_1=\ell$ or $S_2=\ell+1$, and
each record in a bucket has a distinct $SA$ value.

In the \emph{corruption attack}, an adversary has acquired from an
\emph{external source} the $SA$ value $x_i$ of some record $r$ in
the data. $r$ is called a \emph{corrupted record}. Armed with this
knowledge, the adversary will boost the accuracy of inference by
excluding one occurrence of $x_i$ when inferring the sensitive value
of the remaining records that share the same bucket with $r$. To
combat the accuracy boosting, we propose to inject some small number
$\sigma$ of \emph{fake} $SA$ values into each bucket $g$, where a
fake value does not actually belong to any record in the bucket. To
ensure that the adversary cannot distinguish a fake value from a
real value, a fake value must be from the domain of $SA$ and must be
distinct in the bucket.
%
%
Now, for each bucket $g$, the table $QIT$ contains $|g|$ records and
the table $ST$ contains $|g|+\sigma$ distinct $SA$ values, in a
random order. The adversary knows $\sigma$ of these $SA$ values are
fake but does not know which ones.

Suppose now that in a corruption attack, the adversary is able to
corrupt $q$ records in a bucket $g$, where $q\leq |g|$, so
$|g|-q+\sigma$ values remain in $g$, $\sigma$ of which are fake.
Note that $|g|$ and $\sigma$ are constants. Therefore, the more
records the adversary is able to corrupt (i.e., a larger $q$), the
larger the proportion of fake values among the remaining records in
the bucket (i.e., $\frac{\sigma}{|g|-q+\sigma}$) and the more
uncertain the adversary is about whether a remaining value in $g$ is
a real value or a fake value. Even if all but one record in a bucket
are corrupted, the adversary has only $1/(1+\sigma)$ certainty that
a remaining value is a real value. The price to pay for this
additional protection is the distortion by the $\sigma$ fake values
added to each bucket.

The study in \cite{Kifer:2009:APD:1559845.1559861}\cite{Li09}
shows that under unusual circumstances a \emph{negative association}
between a non-sensitive value $z$ and a $SA$ value $x$ may be learnt
from the published data $T^*$, which states that a record having $z$
is less likely to have $x$. Using such negative association, an
adversary could exclude unlikely choices $x$ when inferring the
sensitive value for an individual having the non-sensitive value
$z$. Since this attack shares the same mechanism as the corruption
attack, i.e., by excluding unlikely values, the above solution
proposed for corruption attack can be applied to deter the negative
association attack, with one difference: a fake value should not be
easily excluded for any record using the negative association
knowledge. To ensure this, the publisher can first learn the
negative association from $T^*$ and inject only those fake values
into a bucket that cannot be removed using the learnt negative
association.

\begin{table}[h]
\center
\begin{tabular}{|c|c|} \hline
\textbf{Parameters} & \textbf{Settings}
\\ \hline
Cardinality $|T|$ & 100k, 200k, \textbf{300k}, 400k, 500k  \\ \hline
 $f'_i$-privacy for $x_i$ & $f'_i=min\{1, \theta \times f_i+0.02\}$ \\ \hline
 Privacy coefficient $\theta$ & 2, 4, \textbf{8}, 16, 32  \\ \hline
\hline
 $M$ & $min_i\{\lceil 1/f'_i\rceil\}$ \\ \hline
 $M'$ & 50 \\ \hline
\end{tabular}\caption{Parameter settings} \label{tab:para}
\end{table}

\section{Empirical Studies}\label{evaluation}
\begin{figure}[h]
\centering
\includegraphics[width=9cm,height=2.5cm]{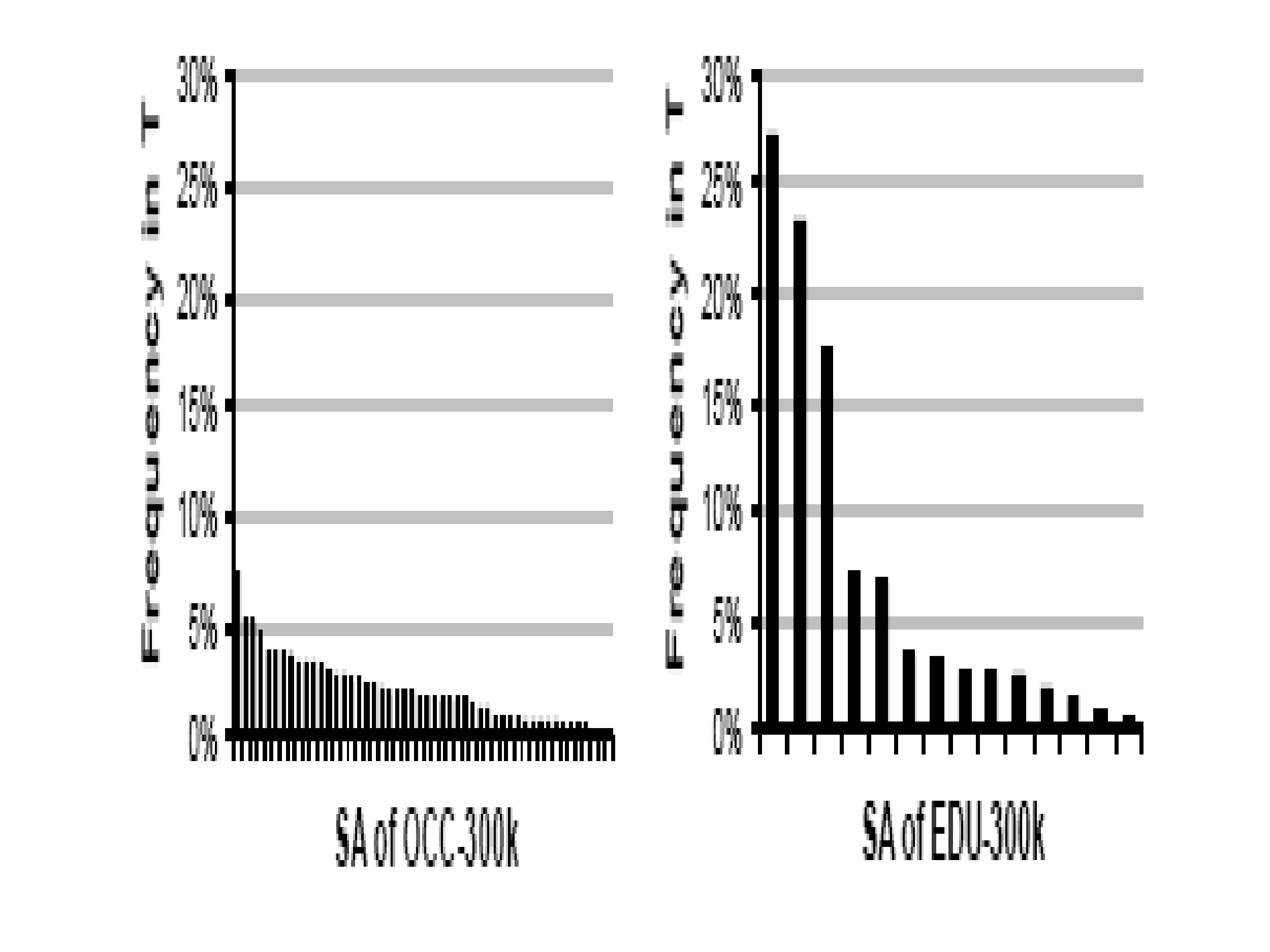}
\caption{Frequency distribution of SA} \label{fig:frequency}
\end{figure}

We evaluate the effectiveness and efficiency of the algorithms
proposed in Section 5. For this purpose, we utilized the real data
set CENSUS containing personal information of 500K American adults.
This data set was previously used in \cite{XT06b}, \cite{LDR05} and
\cite{MKG+006}. Table \ref{tab:census} shows the eight discrete
attributes of the data. Two base tables were generated from CENSUS.
The first table \textit{OCC} has \textit{Occupation} as $SA$ and the
7 remaining attributes as the QI-attributes. The second table
\textit{EDU} has \textit{Education} as $SA$ and the 7 remaining
attributes as the QI-attributes. OCC-n and EDU-n denote the data
sets of OCC and EDU of the cardinality $n$. Figure
\ref{fig:frequency} shows the frequency distribution of $SA$. The
parameters and settings are summarized in Table \ref{tab:para} with
the default setting in bold face.


\begin{small}
\begin{table}[h] \center
\begin{tabular}{|c|c|}
\hline $Attribute$ & Domain Size
\\ \hline \textit{Age} & 76 \\  \hline \textit{Gender} & 2 \\
\hline \textit{Education} & 14 \\ \hline \textit{Marital} & 6 \\
\hline \textit{Race} & 9 \\ \hline \textit{Work-Class} & 10 \\
\hline \textit{Country} & 83 \\ \hline \textit{Occupation} & 50 \\
\hline
\end{tabular}\caption{Statistics of CENSUS} \label{tab:census}
\end{table}
\end{small}

%

We evaluate our algorithms by three criteria: suitability for
handling varied sensitivity, data utility, and scalability.

\subsection{Criterion 1: Handling Variable Sensitivity}\label{l-diversity}
Our first objective is to study the suitability of $F'$-privacy for
handling variable sensitivity and skewed distribution of sensitive
values. For concreteness, we specify $F'$-privacy by $f'_i = min
\{1, \theta\times f_i + 0.02\}$, where $\theta$ is the \emph{privacy
coefficient} chosen from $\{2,4,8,16,32\}$. This specification
models a linear relationship between the sensitivity $f'_i$ and the
frequency $f_i$ for $x_i$. Since $f'_i\geq f_i$ for all $x_i$'s, a
solution satisfying $F'$-privacy always exists (Lemma
\ref{eligibility}). In fact, a solution exists even with the maximum
bucket size constraint $M'=50$.

\begin{figure}[h]
\subfigure[OCC]{
\begin{minipage}[t]{0.45\linewidth}
\centering
\includegraphics[width=4cm,height=3cm]{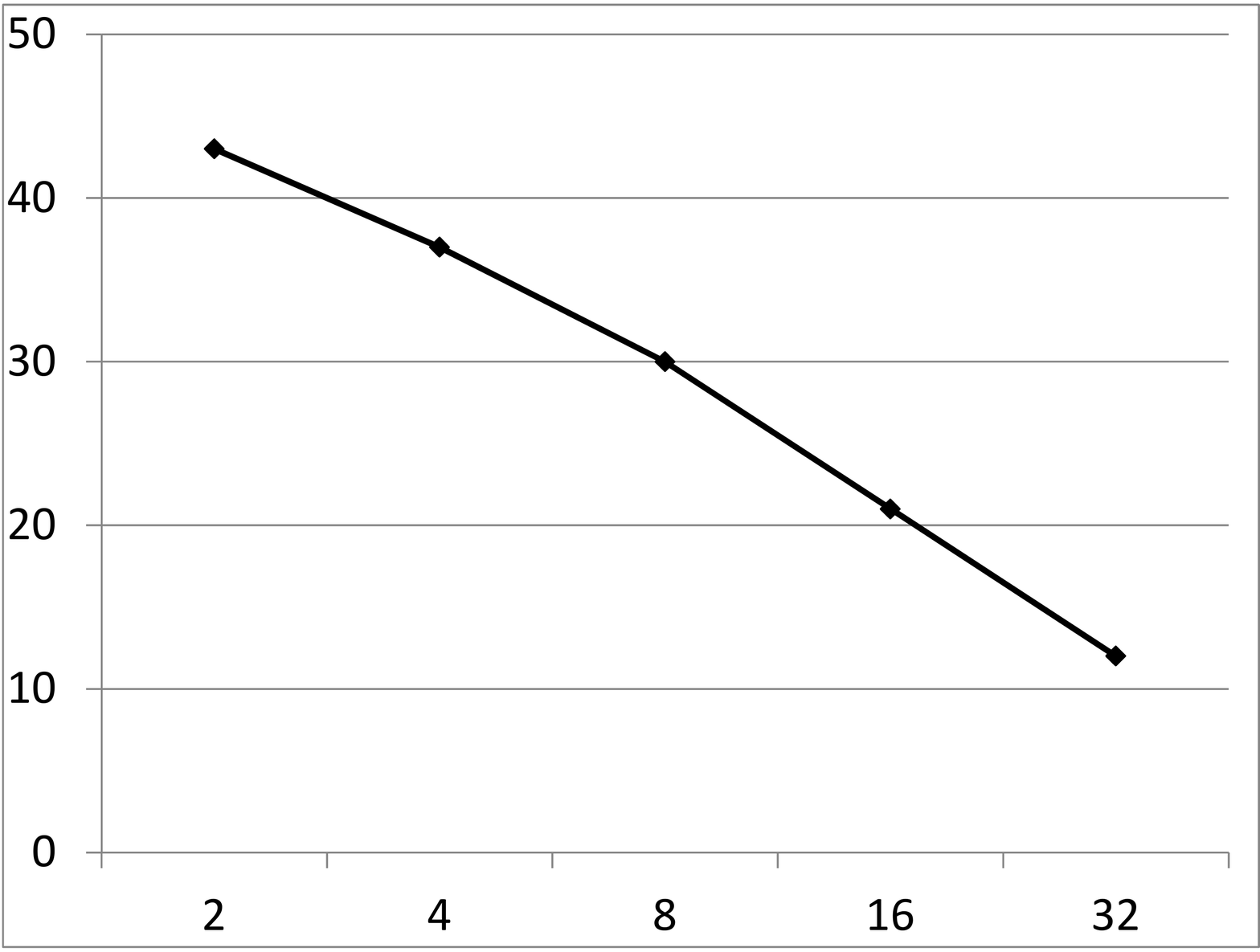}
\end{minipage}}
\hfill \subfigure[EDU]{
\begin{minipage}[t]{0.45\linewidth}
\centering
\includegraphics[width=4cm,height=3cm]{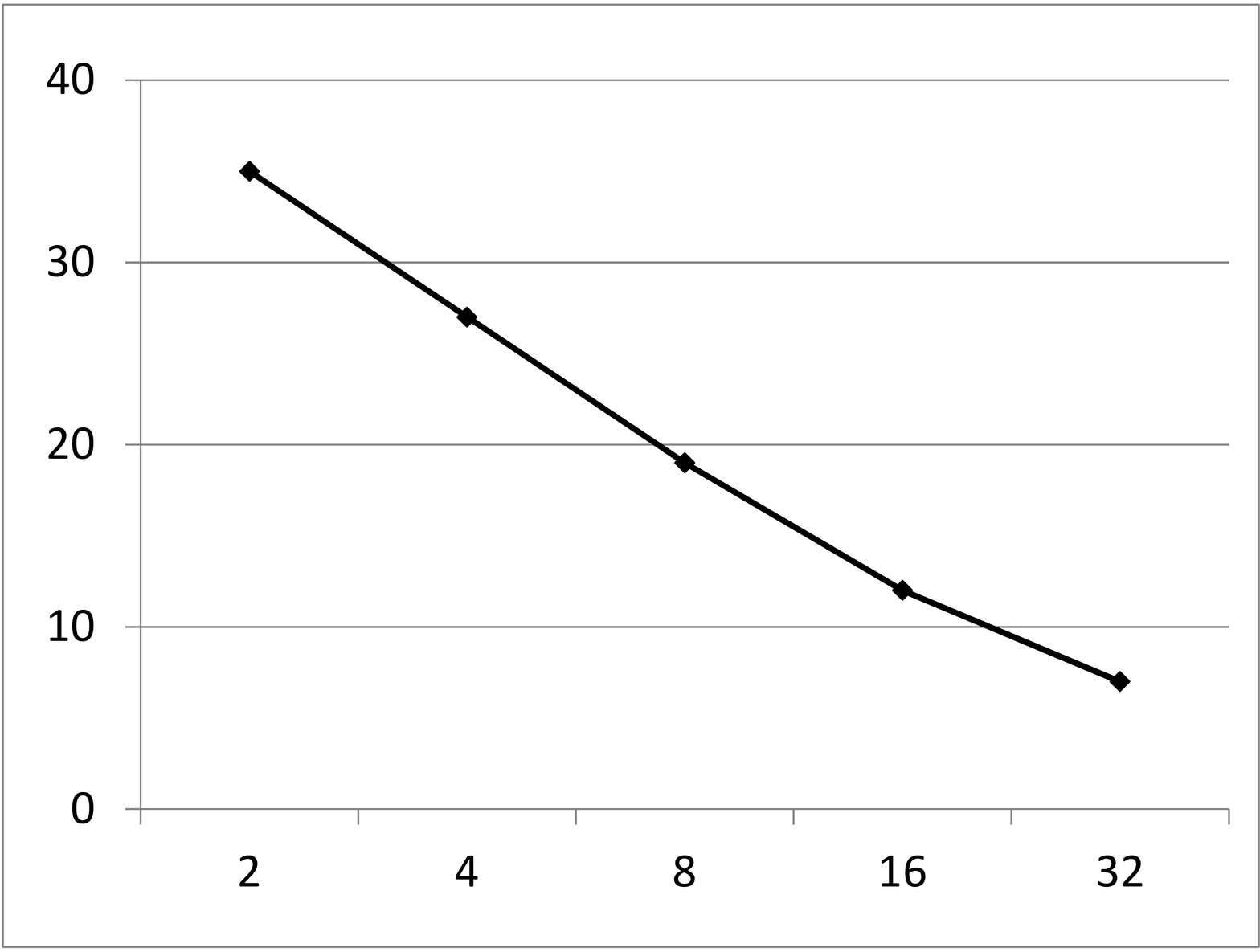}
\end{minipage}}
\caption{The relationship between $\ell$ (y-axis) and privacy
coefficient $\theta$ (x-axis)} \label{lvsa}
\end{figure}

For comparison purposes, we apply $\ell$-diversity to model the
above $F'$-privacy, where $\ell$ is set to $\lceil 1/min_i f'_i
\rceil$ (Remark \ref{remark1}). For the OCC-300K and EDU-300K data
sets, which have the minimum $f_i$ of 0.18\% and 0.44\%,
respectively, Figure \ref{lvsa} plots the relationship between
$\theta$ and $\ell$. Except for $\theta=32$, a rather large $\ell$
is required to enforce $F'$-privacy. As such, the buckets produced
by Anatomy \cite{XT06b} have a large size $\ell$ or $\ell+1$, thus,
a large $Loss$. A large $\ell$ also renders $\ell$-diversity too
restrictive. As discussed in Remark \ref{remark1}, $1/\ell \geq
max_i f_i$ is necessary for having a $\ell$-diversity solution. With
OCC-300K's maximum $f_i$ being 7.5\% and EDU-300K's maximum being
27.3\%, this condition is violated for all $\ell \geq 14$ in the
case of OCC-300K and all $\ell\geq 4$ in the case of EDU-300K, thus,
for most $F'$-privacy considered. This study suggests that
$\ell$-diversity is not suitable for handling sensitive values of
varied sensitivity and skewed distribution.

%

%


\subsection{Criterion 2: Data Utility}
Our second objective is to evaluate the utility of $T^*$. We
consider two utility metrics, \emph{Mean Squared Error (MSE)}
(Definition \ref{MSE}) and \emph{Relative Error (RE)} for count
queries previously used in \cite{XT06b}. We compare
\emph{TwoSizeBucketing}, denoted by ``TwoSize", and
\emph{MultiSizeBucketing}, denoted by ``MultiSize", against two
other methods. (i) \emph{Optimal multi-size bucketing}, denoted by
``Optimal", is the exact solution to the optimal multi-size bucket
setting problem, solved by an integer linear program. ``Optimal"
provides the theoretical lower bound on $Loss$, but it is feasible
only for a small domain size $|SA|$. (ii) \emph{Anatomy}
\cite{XT06b} with $\ell$-diversity being set to $\ell=\lceil 1/min_i
f'_i \rceil$. Except for ``Anatomy", the minimum bucket size $M$ is
set to $min\{\lceil 1/f'_i\rceil\}$ and the maximum bucket size $M'$
is set to 50.

\subsubsection{Mean Squared Error (MSE)}
\begin{figure}[h]
\subfigure[OCC]{
\begin{minipage}[t]{0.45\linewidth}
\centering
\includegraphics[width=4cm,height=3cm]{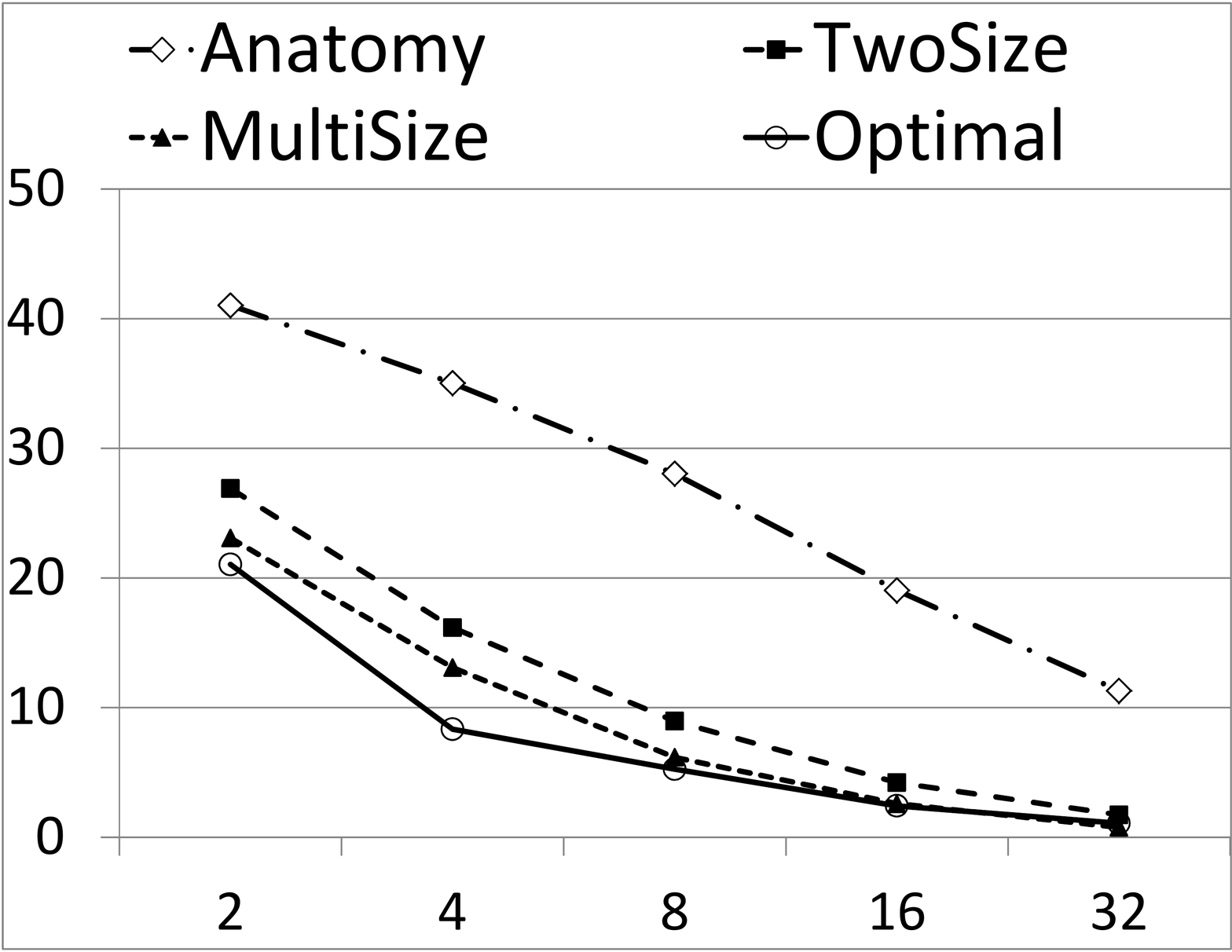}
\end{minipage}}
\hfill
\subfigure[EDU]{
\begin{minipage}[t]{0.45\linewidth}
\centering
\includegraphics[width=4cm,height=3cm]{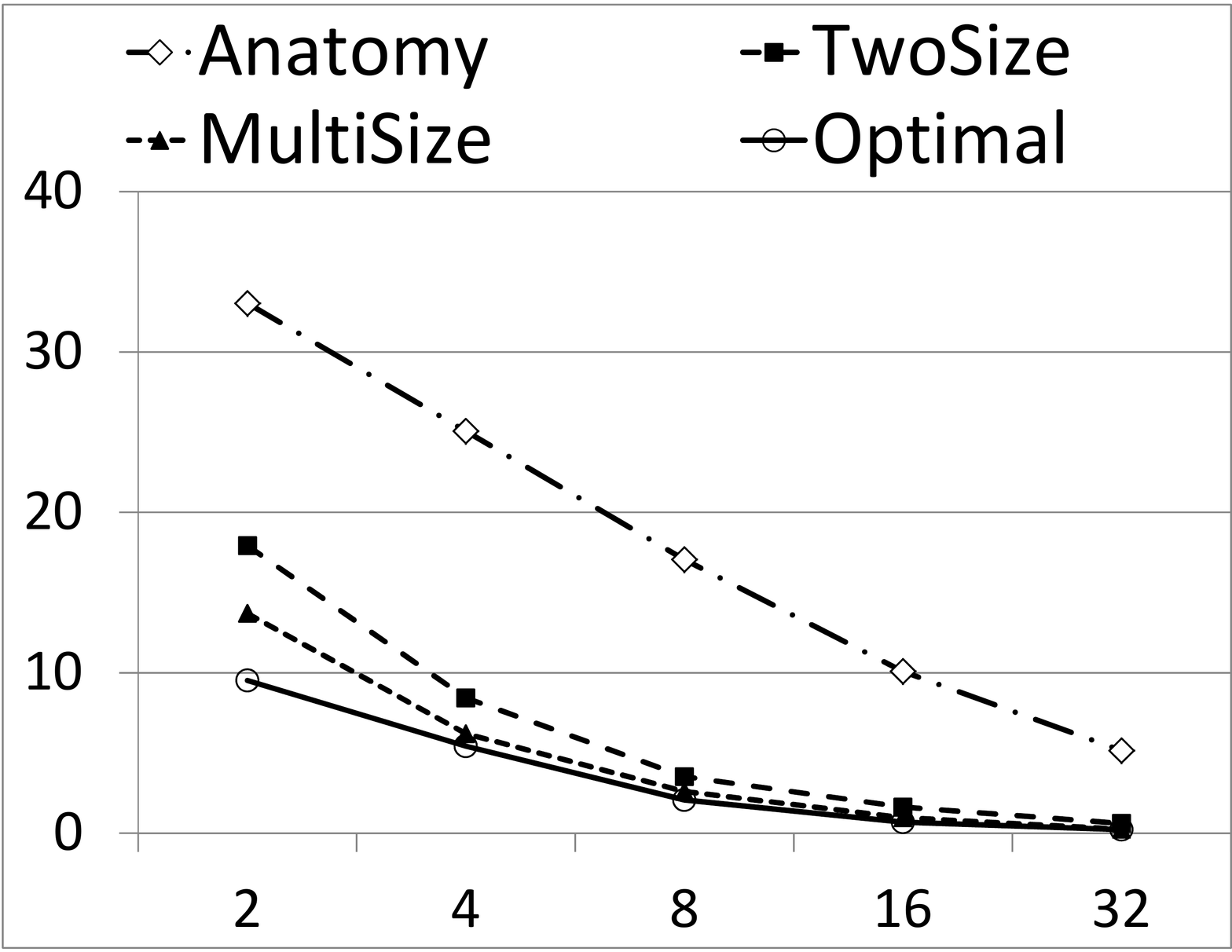}
\end{minipage}}
\caption{MSE (y-axis) vs privacy coefficient $\theta$
(x-axis)}\label{fig:IL}
\end{figure}

Figure \ref{fig:IL} shows $MSE$ vs the privacy coefficient $\theta$
on the default OCC-300K and EDU-300K. The study in Section
\ref{l-diversity} shows that for most $F'$-privacy considered the
corresponding $\ell$-diversity cannot be achieved on the OCC and EDU
data sets. For comparison purposes, we compute the $MSE$ for
``Anatomy" based on the bucket size of $\ell$ or $\ell+1$ while
ignoring the privacy constraint. ``Anatomy" has a significantly
higher $MSE$ than all other methods across all settings of $\theta$
because the bucket sizes $\ell$ and $\ell+1$ are large. ``TwoSize"
has only a slightly higher $MSE$ than ``MultiSize", which has only a
slightly higher $MSE$ than ``Optimal". This study suggests that the
restriction to the two-size bucketing problem causes only a small
loss of optimality and that the heuristic solution is a good
approximation to the optimal solution of the multi-size bucket
setting problem.

%

\subsubsection{Relative Error (RE)}
We adapt \emph{count queries} $Q$ of the form from \cite{XT06b}:\\

\noindent
SELECT COUNT(*) FROM T\\
WHERE $pred(A_1)$ AND ... AND $pred(A_{q_d})$ AND $pred(SA)$\\

\noindent
$A_1,\cdots,A_{q_d}$ are randomly selected QI-attributes. $q_d$ is
the query dimensionality and is randomly selected from
$\{1,\cdots,7\}$ with equal probability, where $7$ is the total
number of $QI$ attributes. For any attribute $A$, $pred(A)$ has the
form

\begin{center}
$A = a_1$ OR ... OR $A = a_b$,
\end{center}

\noindent where $a_i$ is a random value from the domain of $A$. As
in \cite{XT06b}, the value of $b$ depends on the \emph{expected
query selectivity}, which was set to 1\% here. The details can be
found in \cite{XT06b}. The answer $act$ to $Q$ using $T$ is the
number of records in $T$ that satisfy the condition in the WHERE
clause. We created a pool of 5,000 count queries of the above form.
For each query $Q$ in the pool, we compute the estimated answer
$est$ using $T^*$ in the same way as in \cite{XT06b}. The
\emph{relative error (RE)} on $Q$ is defined to be
$RE=|act-est|/act$. We report the average $RE$ over all queries in
the pool.

%

\begin{figure}[h]
\subfigure[OCC]{
\begin{minipage}[t]{0.45\linewidth}
\centering
\includegraphics[width=4cm,height=3cm]{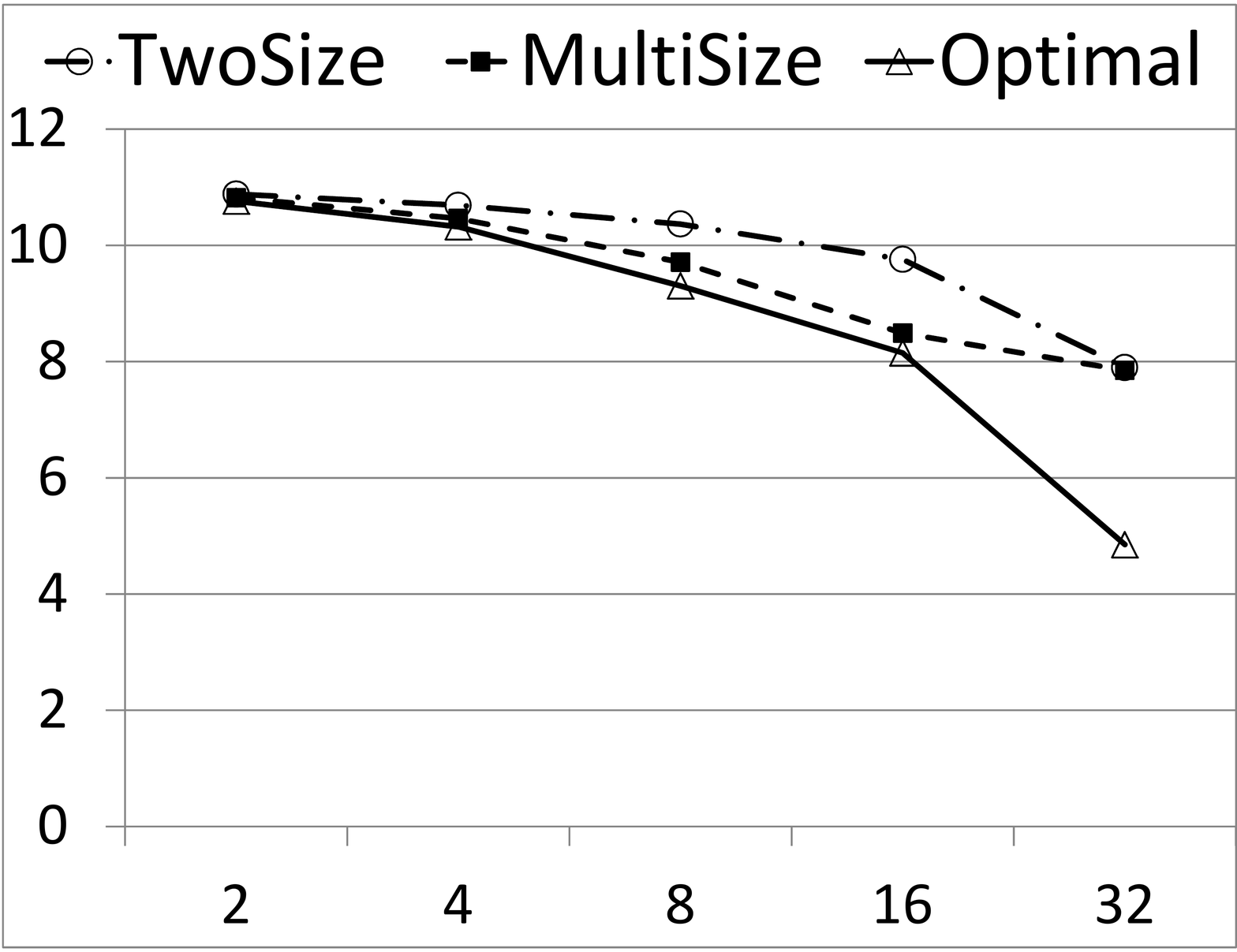}
\end{minipage}}
\hfill
\subfigure[EDU]{
\begin{minipage}[t]{0.45\linewidth}
\centering
\includegraphics[width=4cm,height=3cm]{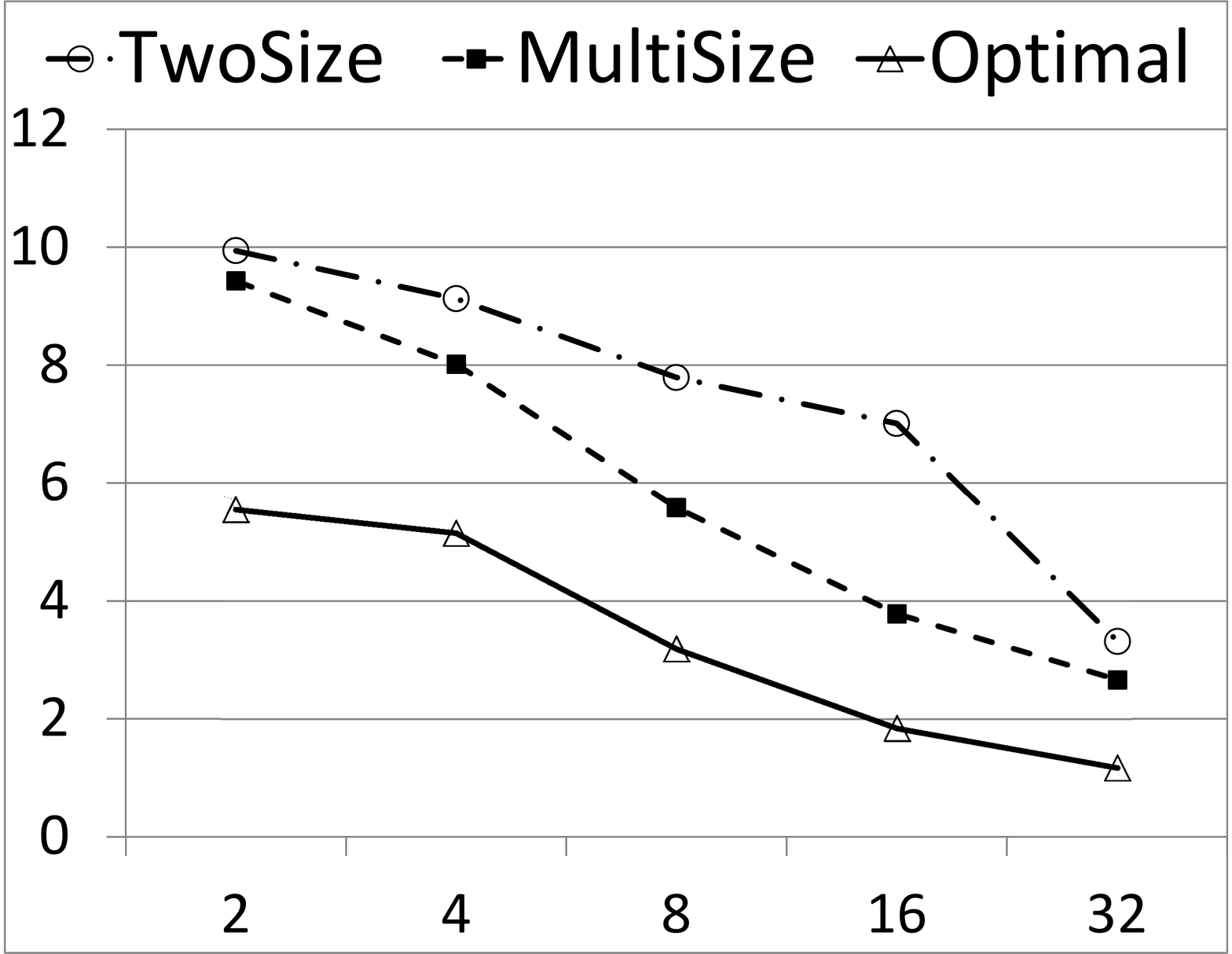}
\end{minipage}}
\caption{Relative Error (\%) (y-axis) vs privacy coefficient
$\theta$ (x-axis)} \label{RelativeErrorvsa}
\end{figure}

Figure \ref{RelativeErrorvsa} shows $RE$ vs the privacy coefficient
$\theta$ on the default OCC-300K and EDU-300K. For the OCC data set,
the maximum $RE$ is slightly over 10\%.
The $RE$'s for ``TwoSize", ``MultiSize",
and ``Optimal" are relatively close to each other, which is
consistent with the earlier finding on similar $MSE$ for these
algorithms. For the EDU data set, all $RE$'s are no more than 10\%.
``MultiSize" improves upon ``TwoSize" by about 2\%, and ``Optimal"
improves upon ``MultiSize" by about 2\%. This study suggests that
the solutions of the optimal two-size bucketing and the heuristic
multi-size bucketing are highly accurate for answering count
queries, with the $RE$ below 10\% for most $F'$-privacy considered.
``Anatomy" was not included since there is no corresponding
$\ell$-diversity solution for most $F'$-privacy considered (see
Section 7.1).


\subsection{Criterion 3: Scalability}
Lastly, we evaluate the scalability for handling large data sets. We
focus on \emph{TwoSizeBucketing} because it is a key component of
\emph{MultiSizeBucketing}. ``No-pruning" refers to the sequential
search of the full list $\Gamma$ without any pruning;
``Loss-pruning" refers to the loss-based pruning in Section 5.1.2;
``Full-pruning" refers to \emph{TwoSizeBucketing} in Section 5.1.3,
which exploits both loss-based pruning and privacy-based pruning.
``Optimal" refers to the integer linear program solution to the
two-size bucketing problem. We study the \emph{Runtime} with respect
to the cardinality $|T|$ and the domain size $|SA|$. The default
privacy coefficient setting $\theta=8$ is used. All algorithms were
implemented in C++ and run on a Windows 64 bits Platform with CPU of
2.53 GHz and memory size of 12GB. Each algorithm was run 100 times
and the average time is reported here.

%

\subsubsection{Scalability with $|T|$}

%

\begin{figure}[h]
\subfigure[OCC]{
\begin{minipage}[t]{0.45\linewidth}
\centering
\includegraphics[width=4cm,height=3cm]{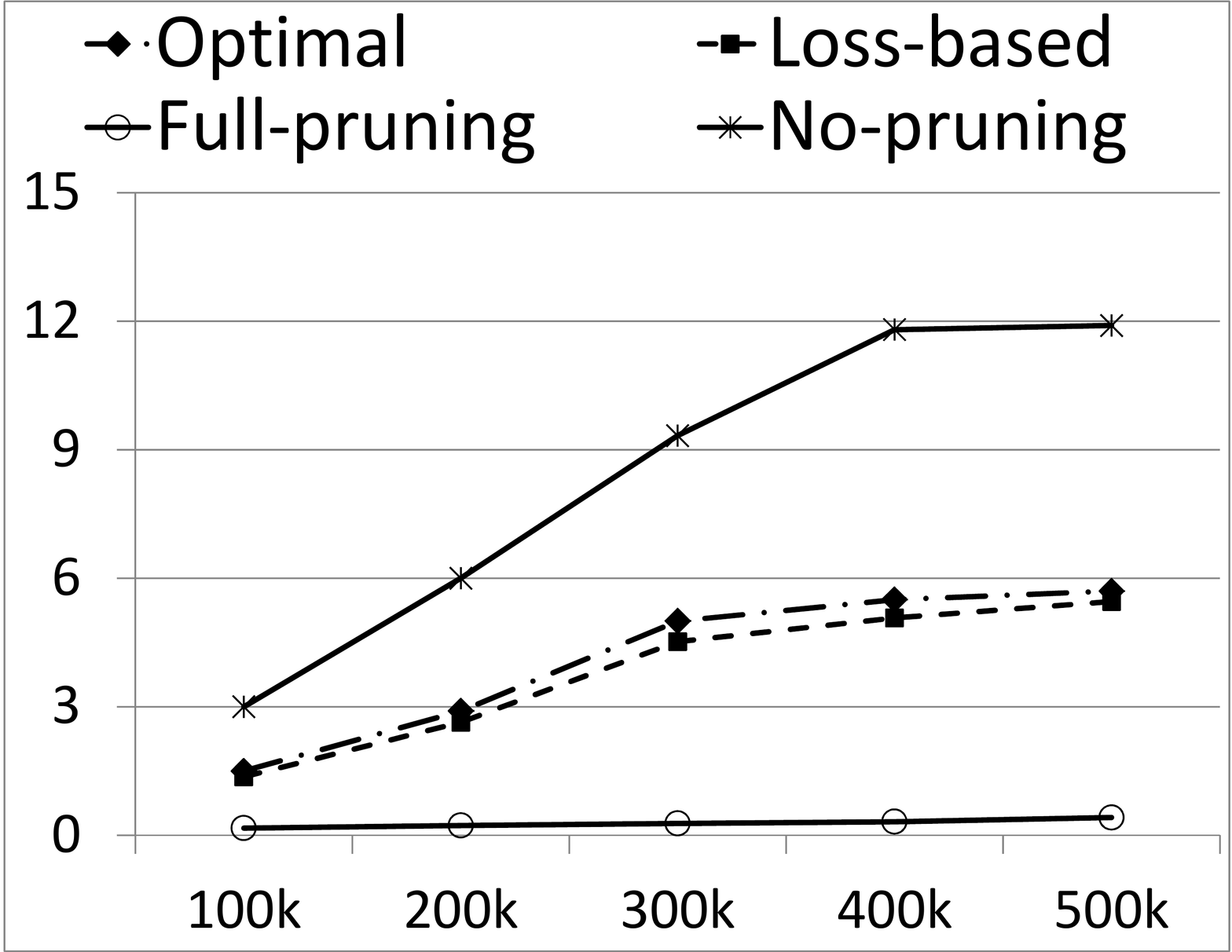}
\end{minipage}}
\hfill
\subfigure[EDU]{
\begin{minipage}[t]{0.45\linewidth}
\centering
\includegraphics[width=4cm,height=3cm]{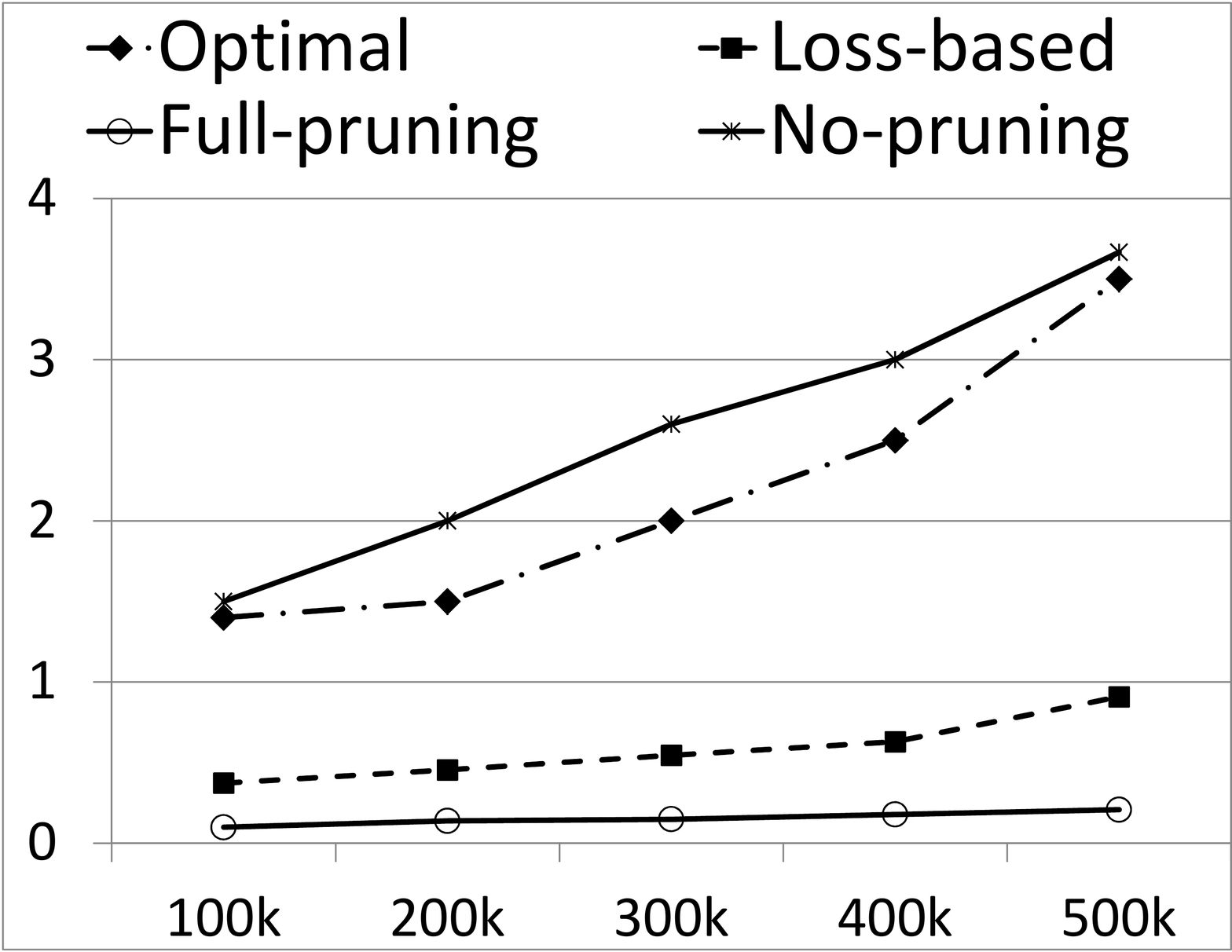}
\end{minipage}}
\caption{Runtime (seconds) (y-axis) vs cardinality $|T|$ (x-axis)}
\label{Runningtime-n}

\subfigure[OCC]{
\begin{minipage}[t]{0.45\linewidth}
\centering
\includegraphics[width=4cm,height=3cm]{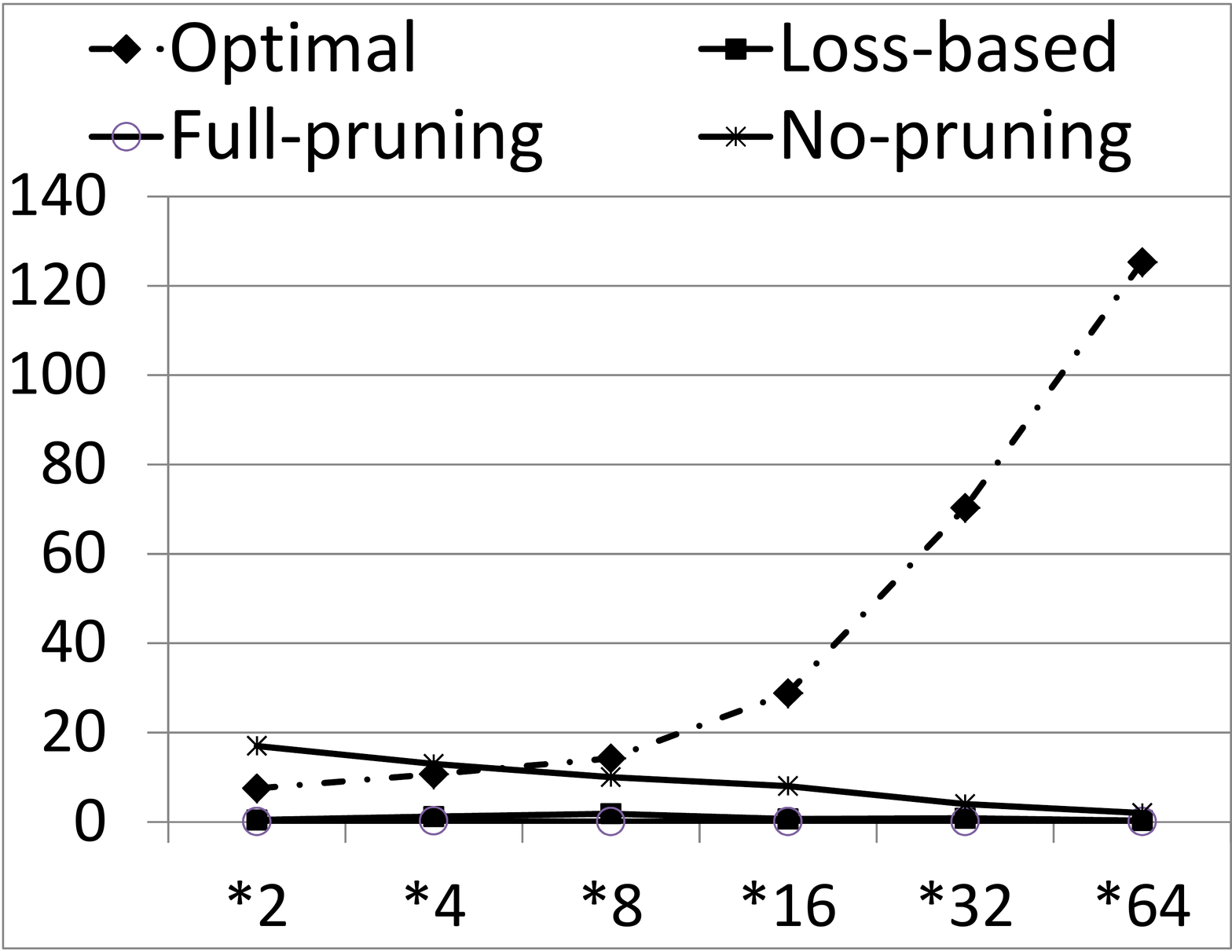}
\end{minipage}}
\hfill \subfigure[EDU]{
\begin{minipage}[t]{0.45\linewidth}
\centering
\includegraphics[width=4cm,height=3cm]{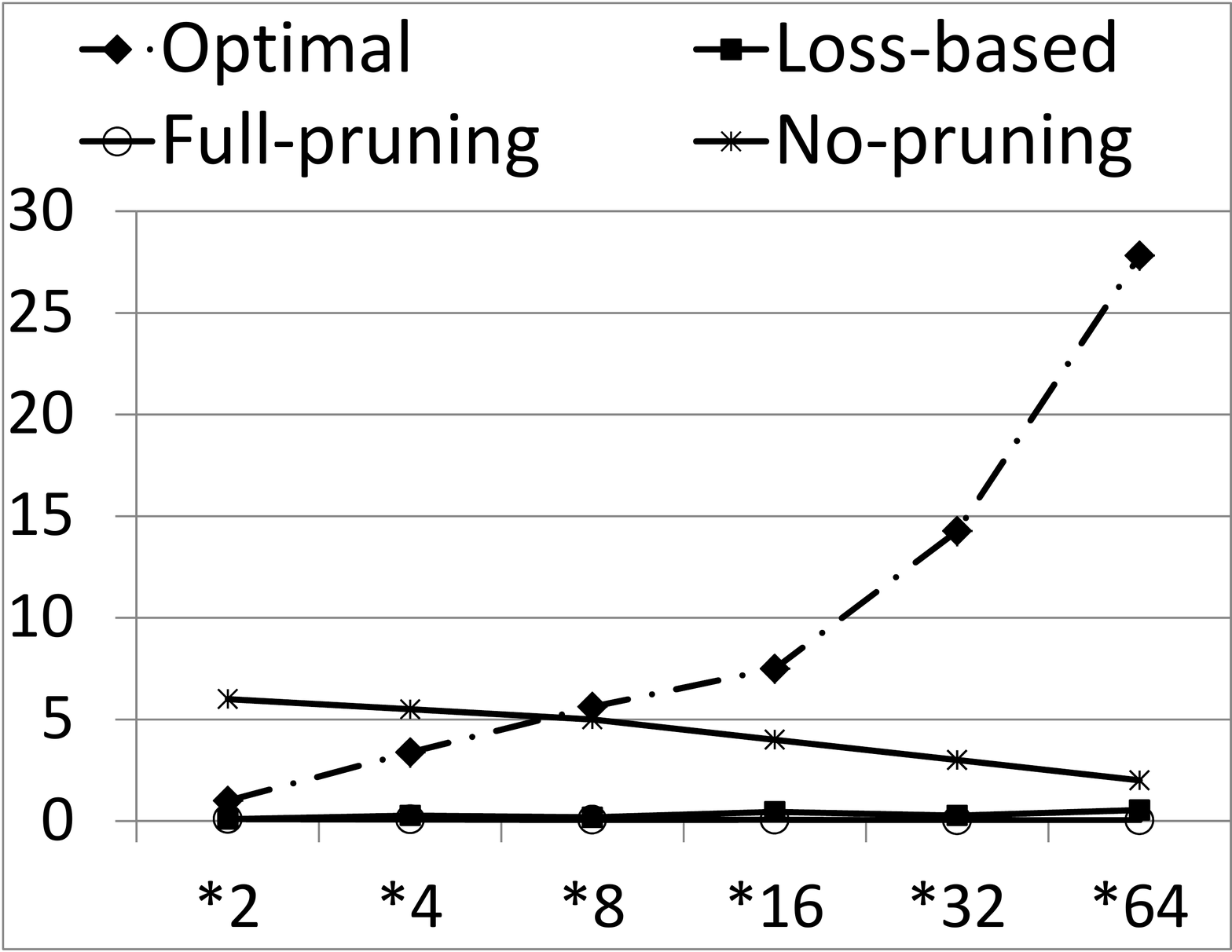}
\end{minipage}}
\caption{Runtime (seconds) (y-axis) vs scale-up factor $\gamma$ for
$|SA|$ (x-axis)} \label{Runningtimevstimes}
\end{figure}

Figure \ref{Runningtime-n} shows $Runtime$ vs the cardinality $|T|$.
``Full-pruning" takes the least time and ``No-pruning" takes the
most time. ``Loss-pruning" significantly reduces the time compared
to ``No-pruning", but has an increasing trend in $Runtime$ as $|T|$
increases because of the sequential search of the first valid pair
in the list $\Gamma'$. In contrast, a larger $|T|$ does not affect
``Full-pruning" much because ``"Full-pruning" locates the first
valid pair by a binary search over $\Gamma'$. ``Optimal" takes less
time than ``No-pruning" because the domain size $|SA|$ is relatively
small. The next experiment shows that the comparison is reversed for
a large domain size $|SA|$.


\subsubsection{Scalability with $|SA|$}
We scale up $|SA|$ for OCC-500K and EDU-500K by a factor $\gamma$,
where $\gamma$ is ranged over 2, 4, 8, 16, 32 and 64. Assume that
the domain of $SA$ has the form $\{0,1,\cdots,m-1\}$. For each
record $t$ in $T$, we replace $t[SA]$ in $t$ with the value $\gamma
\times t[SA]+r$, where $r$ is an integer selected randomly from the
range $[0,\gamma-1]$ with equal probability.
Thus the new domain of $SA$ has the size $m\times \gamma$. Figure
\ref{Runningtimevstimes} shows $Runtime$ vs the scale-up factor
$\gamma$. As $\gamma$ increases, $Runtime$ of ``Optimal" increases
quickly because the integer linear programming is exponential in the
domain size $|SA|$. $Runtime$ of the other algorithms increases
little because the complexity of these algorithms is linear in the
domain size $|SA|$. Interestingly, as $|SA|$ increases, $Runtime$ of
``No-pruning" decreases. A close look reveals that when there are
more $SA$ values, $f_i$ and $f'_i$ become smaller and the minimum
bucket size $M$ becomes larger, which leads to a short $\Gamma$
list. A shorter $\Gamma$ list benefits most the sequential search
based ``No-pruning".

In summary, we showed that the proposed methods can better handle
sensitive values of varied sensitivity and skewed distribution,
therefore, retain more information in the data, and the solution is
scalable for large data sets.


\section{Conclusion}
Although differential privacy has many nice properties, it does not
address the concern of inferential privacy, which arises due to the
wide use of statistical inferences in advanced applications. On the
other hand, previous approaches to inferential privacy suffered from
major limitations, namely, lack of flexibility in handling varied
sensitivity, poor utility, and vulnerability to auxiliary
information. This paper developed a novel solution to overcome these
limitations. Extensive experimental results confirmed the
suitability of the proposed solution for handling sensitive values
of varied sensitivity and skewed distribution.

%
%

\label{conclusion}
\bibliography{minmax5reference}
\end{document}